\documentclass[modern, 12pt]{aastex62}
\usepackage{longtable}
\usepackage[usestackEOL]{stackengine}
\usepackage{graphicx}
\usepackage{xcolor}
\usepackage{natbib}
\usepackage{amssymb,amsmath}
\NewPageAfterKeywords

\shorttitle{Sifting Through the Static}
\shortauthors{Smotherman et al.}

\begin{document}

\correspondingauthor{Hayden Smotherman}
\email{smotherh@uw.edu}

\title{Sifting Through the Static: Moving Object Detection in Difference Images}

\author{Hayden Smotherman} \affil{Department of Astronomy, University of Washington, Seattle, WA, 98195, USA}
\author{Andrew J. Connolly} \affil{Department of Astronomy, University of Washington, Seattle, WA, 98195, USA}
\author{J. Bryce Kalmbach} \affil{Department of Astronomy, University of Washington, Seattle, WA, 98195, USA}
\author{Stephen K. N. Portillo} \affil{Department of Astronomy, University of Washington, Seattle, WA, 98195, USA}
\author{Dino Bektesevic} \affil{Department of Astronomy, University of Washington, Seattle, WA, 98195, USA}
\author{Siegfried Eggl} \affil{Department of Astronomy, University of Washington, Seattle, WA, 98195, USA} \affil{Department of Aerospace Engineering, University of Illinois, Urbana, IL, 61801, USA}
\author{Mario Juric} \affil{Department of Astronomy, University of Washington, Seattle, WA, 98195, USA}
\author{Joachim Moeyens} \affil{Department of Astronomy, University of Washington, Seattle, WA, 98195, USA}
\author{Peter J. Whidden} \affil{Department of Astronomy, University of Washington, Seattle, WA, 98195, USA}

%\keywords{methods: data analysis, techniques: image processing, difference imaging, minor planets, asteroids: general, Kuiper belt: general}

\begin{abstract}

Trans-Neptunian Objects (TNOs) provide a window into the history of the Solar System, but they can be challenging to observe due to their distance from the Sun and relatively low brightness. Here we report the detection of 75 moving objects that we could not link to any other known objects, the faintest of which has a \textit{VR} magnitude of $25.02 \pm 0.93$ using the KBMOD platform. We recover an additional 24 sources with previously-known orbits. We place constraints on the barycentric distance, inclination, and longitude of ascending node of these objects. The unidentified objects have a median barycentric distance of 41.28 au, placing them in the outer Solar System. The observed inclination and magnitude distribution of all detected objects is consistent with previously published KBO distributions. We describe extensions to KBMOD, including a robust percentile-based lightcurve filter, an in-line graphics processing unit (GPU) filter, new coadded stamp generation, and a convolutional neural network (CNN) stamp filter, which allow KBMOD to take advantage of difference images. These enchancements mark a significant improvement in the readiness of KBMOD for deployment on future big data surveys such as LSST.

\end{abstract}

\section{Introduction} \label{sec:intro}

Small bodies are the final frontier in the study of flux-limited populations in the Solar System. While these objects are primarily very small and often very distant, they are nevertheless critical to our understanding of the formation of the Solar System. For example, Trans-Neptunian Objects (TNOs) contain dynamically-unperturbed relics from the formation of the Solar System \citep{Luu+Jewitt2002}. They provide a window into the early history of the Solar System and enable tests of planetary formation and migration hypotheses. The Nice model \citep{NiceModel} suggests that all the giant planets formed well-interior to 20 au and migrated outwards due to interactions with planetesimals. Better knowledge of the dynamical populations of TNOs would enable tests of additional and alternative hypotheses regarding the dynamical history of the Solar System, such as the smooth migration of Neptune \citep{Hahn+2005, NesvornyDavid2015JNCE, MorbidelliAlessandro2019Kbfa}, a stellar flyby \citep{Kenyon+2004}, or rogue planetary embryos \citep{Gladman+2006}. Improving our understanding of TNO size and orbital distributions, especially at the low-mass end where they are more poorly constrained, will be critical for our understanding of these and other hypotheses.

Well beyond the edge of the Classical Kuiper Belt lies the Oort Cloud. Most famously, the Inner Oort Cloud includes the object Sedna, which is thought to be a member of a larger population of sednoids \citep{Brown_2004}. Sedna is an Inner Oort Cloud object with a perihelion of $76\pm4$ au, but a semimajor axis of $480\pm40$ au. Therefore, it spends well over $90\%$ of the time on its orbit beyond the detection limit of the survey that discovered it. According to the Minor Planet Center (MPC) database for TNOs, centaurs, and scattered disk objects (SDO), Sedna has the second largest perihelion (after 2012 VP113) of any detected Solar System object. It represents one of the only currently-observable links to the Oort Cloud, a region that contains a wealth of information about the history of the Solar System. If we could increase the number of known sednoids and other Oort Cloud objects, they would provide observational constraints on the formation environment of the Sun \citep{BRASSER200659} and the Sun's dynamic history in the Milky Way after leaving its formation environment \citep{KAIB2011491}.

New and upcoming approaches to survey astronomy provide exciting opportunities for the study of these populations. For example, the upcoming Legacy Survey of Space and Time \citep[LSST; lsst.org;][]{Ivezic_2019} expects to survey over 18,000 square degrees of the sky 825 times over a period of 10 years, generating about 20 TB of data every 24 hours. 

LSST plans to detect Solar System objects from individual images, with a single-visit limiting magnitude in the r band of 24.7, and link these detections to measure orbits. Current projections \citep{lsstsciencebook} show that LSST is expected to detect about 40,000 TNOs, which is by itself a large increase over the currently-known 4077 Centaurs, KBOs, and SDOs (MPC). However, if we could coadd the images to increase the signal-to-noise ratio (SNR) of the Solar System detections, then we could recover significantly more objects. Following the formula $\Delta m = (5/2) \log{\sqrt{N}}$, coadding just three months of LSST data would increase the limiting magnitude in the $r$ band from 24.7 to 26.1. This increase in depth means LSST would detect $\sim 8.0$ times more TNOs compared to a single image, assuming the single power-law $r$ band KBO distribution of \citet{FRASER2008827}. Instead of 40,000 TNOs, we could detect $\sim 320,000$ TNOs. If we could coadd a year of LSST data, this increases to over 520,000 new TNOs detected (given our simplified assumptions). None of this requires any more data than LSST will already acquire.

Coadding moving objects poses unique challenges compared to coadding stars. Because stars move very slowly compared to most survey cadences, coaddition of a stack of aligned images usually increases the limiting magnitude for stars compared to single images. Solar System objects, however, generally move at on-sky velocities of $>1''\ \mathrm{hr^{-1}}$, due to both the proper motion of the objects and the reflex motion caused by the Earth's orbit. This means that traditional image coaddition typically does not increase the number of detectable moving objects. Known moving objects may be tracked and aligned along their orbits to improve the quality of the detection, but to use image coaddition to detect new objects with unknown orbits, another approach is required.

The Kernel-Based Moving Object Detection \citep[KBMOD;][]{kbmod} algorithm takes a time series of images of the same RA and Dec, uses a ``track before detect" (TBD) approach to account for the potential motion of objects on an image,  and then coadds the shifted images (increasing the SNR of objects with the candidate trajectory). To sample all possible orbital parameters requires searching billions of candidate trajectories even within the footprint of a single charge-coupled device (CCD). Consequently, current implementations of TBD have generally been restricted  to narrow-field surveys \citep{Bernstein_2004}. KBMOD addresses this by using GPU-accelerated computing to search over a wide range of trajectories for a stack of CCDs in of order 10 minutes.

In this paper, we present a number of algorithmic improvements to KBMOD that allow us to search for moving objects in difference images. We use the Dark Energy Camera (DECam) NEO Data Survey to validate our improvements. This is a larger survey with a longer and more irregular cadence than KBMOD has been applied to in the past. Successfully running on difference images and a more complicated survey shows that KBMOD is beginning to be applicable at the scale needed for upcoming big data surveys like LSST. In Section \ref{sec:data}, we discuss the DECam NEO Data Survey and the processing we applied to it using the LSST Software Stack. In Section \ref{sec:tech}, we discuss the KBMOD algorithm and present recent improvements. In Section \ref{sec:results}, we discuss the results from our analysis, including the detection of unidentified outer Solar System objects. We discuss current limitations and future improvements in Section \ref{sec:discuss}.
\section{Data} \label{sec:data}

\subsection{The DECam NEO Survey Data} \label{sec:neodata}

The DECam NEO Data Survey covered an area on the sky of greater than 2000 square degrees. The $\sim$6.7 TB data set from the DECam NEO Data Survey (PI Lori Allen) uses the Dark Energy Camera on the 4m Blanco telescope at the Cerro Tololo Inter-American Observatory (CTIO) \citep{Flaugher_DECam_Instrument}. The DECam NEO Data Survey consists of 32 nights of data.  In the first 10-night observing run in 2014, \citet{Trilling2017} found 235 unique NEOs.

Each individual image taken by DECam is a composite of 62 2K x 4K science CCDs, with a fill factor of 0.8 \citep{HERNER2020100425}. Each CCD image covers an area of $\sim$0.04 square degrees with a pixel scale of 0.27 arcseconds. This results in a total field of view for DECam of about 3 square degrees. The CCDs are $250\ \mathrm{\mu m}$ thick fully depleted devices, with a peak quantum efficiency above $85\%$ at $\sim6500$\AA \citep{Flaugher_DECam_Instrument}. Gaps between CCDs are between 153 pixels (columns) and 201 pixels (rows). Observations for this data set were taken in the \textit{VR} filter, a broad optical filter extending from 500 to 760 nm.

We separate this data set into 782 pointing groups based on RA and Dec. CCD 01 and 61 had no data in our images, leading to a set of 60 CCDs per pointing group. We define a pointing group as a set of DECam exposures within $25''$ of a common RA and Dec and define a pointing as an individual DECam exposure (i.e. a set of 60 CCDs) in a pointing group. Most pointing groups contain between 5 and 25 pointings. Pointing groups characteristically have 5 pointings per night, with all data taken over nearly-consecutive nights. The intra-night pointings are taken about five minutes apart for a total intra-night timespan of approximately 25 minutes.

24 pointing groups had a high stellar number density, with more than 10000 sources detected in a CCD. When astrometrically calibrating these images (see Section \ref{sec:dataprocess}), these pointing groups exceeded the memory limits of the available computational resources and were therefore excluded. The current limitations regarding the processing of dense fields with LSST Science Pipelines are described in \citet{sullivan_ian_2021_5172677}. Detectability of moving objects with KBMOD, however, is driven strongly by the quality of the difference images.

372 pointing groups contained data from at least four unique survey nights. Because of the short intra-night image cadence, which can cause slow-moving objects to exhibit minimal motion within a night, we only search over pointing groups with at least four unique survey nights. This ensures that any given KBMOD trajectory will search a sufficiently-large number of unique on-sky positions, thereby reducing the probability of linking of static objects.

In order to comply with computational limitations, we selected 43 pointing groups from the set of 372 pointing groups, focusing our research on higher-quality data. These 43 pointing groups have a total effective search area of approximately 132 square degrees. We refer to these 43 pointing groups as the ``search sample''. This down select from 372 pointing groups was as follows. 40 pointing groups existed where all pointings in the pointing group had a maximum seeing full width at half maximum (FWHM) of $1.25''$. These 40 pointing groups make up the bulk of the search sample. There were an additional 12 pointing groups that had over 20 total pointings, but with only 20 pointings with seeing $< 1.25''$. These pointing groups returned a greater number of erroneous candidate trajectories that required by-eye rejection. This is possibly due to the inclusion of poor-seeing images in the image differencing template (see Section \ref{sec:dataprocess}). Due to computational limitations, we elected to run KBMOD on only 3 of these pointing groups, focusing our GPU resources on the 40 pointing groups where all 20 pointings had the required seeing limits. These 3 pointing groups make up the remainder of the search sample.

\subsection{Processing the DECam Data} \label{sec:dataprocess}

The raw DECam images were processed by the DECam Community Pipeline \citep{Valdes2014} resulting in a set of InstCal PROCTYPE images, as defined in the NOAO Data Handbook \citep{NOAO+handbook}. These images are bias and linearity corrected, flat-fielded, and sky-subtracted by the community pipeline. Data quality masks and inverse variance arrays were provided. We downloaded the compressed InstCal data from the NOAO Data Archive between July and November of 2017.

Prior to running the KBMOD pipeline, we first astrometrically calibrate the images in all 782 pointing groups. This was undertaken using the LSST Science Pipelines Software \citep{LSST_DM}. Sources were detected in the individual images. Sources with a $\mathrm{SNR}>=40$ were matched to the data from the GAIA Data Release 1 (DR1). The median astrometric scatter for the sources used to fit the CCD world coordinate systems (WCS) was 25 mas; 373 CCDs had an astrometric scatter worse than 100 mas. The median number of sources detected per CCD was 3575.

\begin{figure}
    \centering
    \includegraphics[width=0.9\textwidth,keepaspectratio]{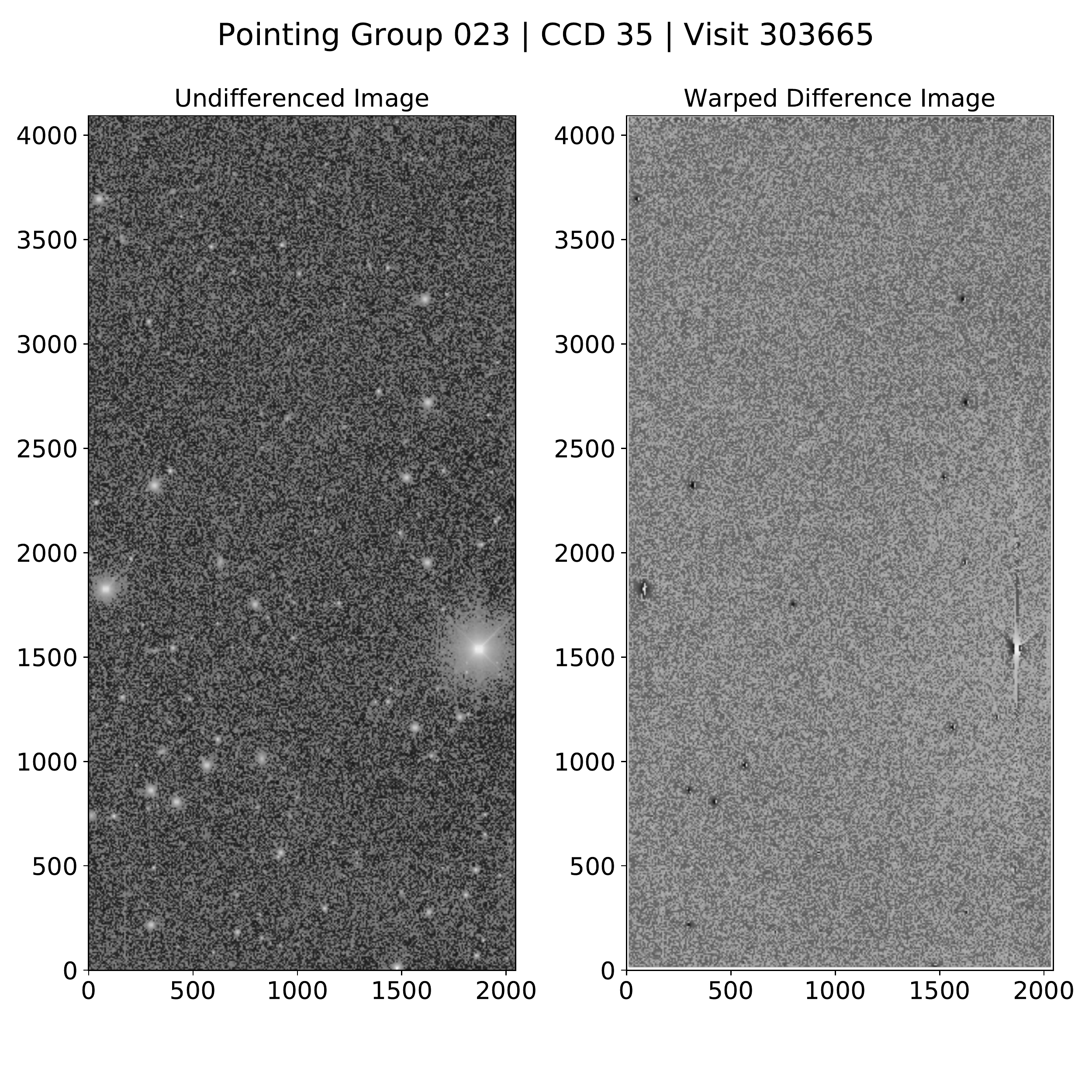}
    \caption{Single pointing  (pointing group 011, CCD 29, visit 303605) before (left) and after (right) image differencing. Similar to DS9, we applied an arcsinh filter to the pixels in this example in order to better show objects in each image.} \label{fig:diffexp}
\end{figure}

\begin{figure}
    \centering
    \includegraphics[width=0.9\textwidth,keepaspectratio]{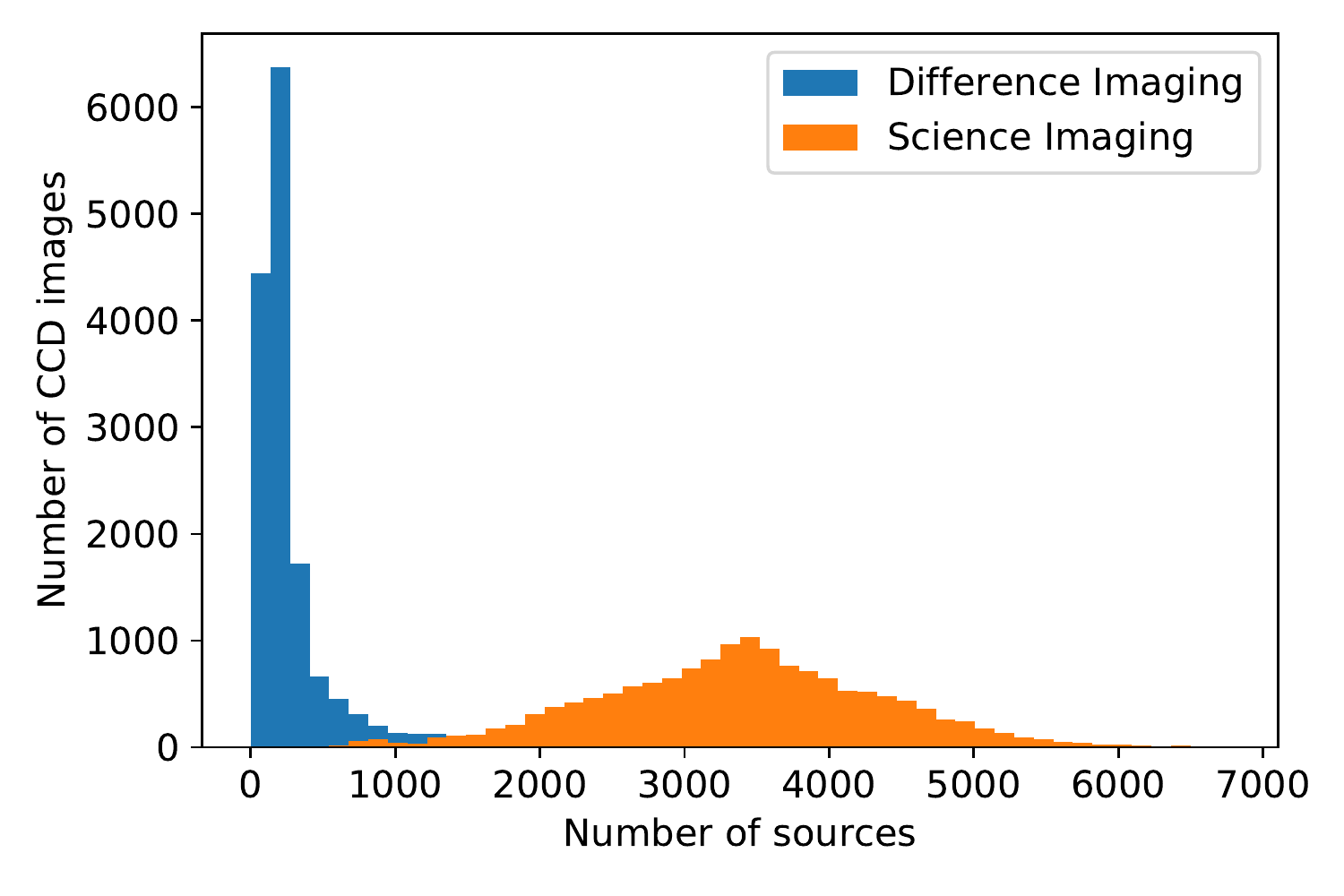}
    \caption{Number of sources per CCD image for each visit in 10 pointing groups (pointing group 091 to 100). The median number of sources per science image (orange) is 3396 per CCD image. The median number of sources per difference image (blue) is 180 per CCD image. Differencing the science images therefore reduces the number of static sources in the image by a factor of about 18.} \label{fig:sourcecomp}
\end{figure}

As a followup to \cite{kbmod}, we use image differencing to remove non-variable and non-moving sources within an image (as opposed to just masking the sources). We used the LSST Stack to difference the images in the pointing groups using a method based on \citet{Alard1998}. For the DECam NEO Survey data, we difference each pointing against a coadded template.  Given the short intra-night time separation between images of a given pointing group, objects moving slower than of order $1"\ \mathrm{hr}^{-1}$ will not move a full psf width over a single night. We therefore separate a pointing group into two approximately equal groups such that each image in the first group will be separated in time from each image in the second group by at least twelve hours. A coadded template was independently generated from each group and used to difference the opposite group. Because our minimum search velocity is $\geq 92$ pixels per day ($\geq 24"$ per day), this guarantees that objects of interest will be much greater than one PSF away from where they were in the coadded template. This means that pointings in the middle of a pointing group---with respect to time --- will have the shortest image differencing baseline, and will therefore set a theoretical limit on the slowest-moving objects we can detect.

In order to difference the science images against the coadded template \citep{Alard1998,2017ApJ...836..188Z,2017ApJ...836..187Z,2016ApJ...830...27Z}, we need to find a convolution kernel $K$ such that for a science image $I(x)$ and a coadded template $\Phi(x)$, $I(x) = K \otimes \Phi(x)$. Following the approach of \citet{Alard1998}, we separated the template into local spatial cells of 128x128 pixels. We detected sources in both images, and grouped them into the spatial cells. Stamps of these sources were created with sizes between 21x21 pixels and 35x35 pixels, depending on the FWHM of the source. Stamps in each cell were used to find the local spatially-invariant convolution kernel solutions of each stamp. The local convolutional kernel was modelled as a set of Gaussian functions multiplied with a polynomial. The coefficients of the kernel were then found by solving a least-squares problem. One source (and thus one stamp) was selected for each grid cell based on the clipped mean of all the kernel solutions in the cell. This gave the local convolution kernel for that cell. Chebyshev polynomials of the first kind were fit to the local kernel coefficients in order to determine a model for spatially-variant global convolution kernel coefficients. This global kernel was then used to match the PSF of the coadded template to that of the science image. We matched the template to the science image, rather than the other way around, because the template has less noise and the convolution correlates noise. The two images were then subtracted. Finally, a decorrelation algorithm was run to remove the correlation in the noise of the difference image. After differencing the image, we warp all images in a pointing group to the sky plane of the first pointing in the pointing group. This ensures that a pixel in one pointing will correspond to the same RA and Dec as that of the same pixel in another pointing.

As an example, Figure \ref{fig:diffexp} shows pointing group 023, CCD 35, visit 303665 before and after image differencing and warping. The final image size for the KBMOD image is set by the intersection of the image and the template that is subtracted. Slight misalignments of the pointings in a pointing group may reduce the final image sizes. All pointing groups were, however, aligned to within $50$ arcsec in RA and Dec, with all but 28 pointing groups aligned to better than $25$ arcsec in both RA and Dec. Therefore the reduction in image area was minimal.

The asteroid search was run for each aligned stack of DECam CCDs independently; we did not search trajectories across CCD boundaries. The effective area on which we are able to search for moving objects is, therefore, about 0.04 square degrees. In other words, a necessary requirement for the detection of a moving object with the KBMOD algorithm is that the object stays within the field of view of an individual CCD for at least two pointings. In practice, we require that an object stay in the field for at least 3 nights (typically 15 pointings). This means that an object must move slower than about $15"\ \mathrm{hr}^{-1}$ to be detected by KBMOD.

\section{Techniques} \label{sec:tech}
KBMOD generates images of likelihood ($\Psi_i$) and variance ($\Phi_i$) from a series of CCD images as described in \citet{kbmod}. Assuming a Gaussian likelihood function, a stack of $\Psi_i$ and $\Phi_i$ images can then be shifted along a potential asteroid trajectory and summed in order to get the coadded likelihood of a detection ($\Psi_{coadd} = \sum_i \Psi_i$ and $\Phi_{coadd} = \sum_i \Phi_i$). See \citet{2018AJ....155..169O} for the optimal approach for source detection with Poisson noise. We define a SNR $\nu$ for a detection such that $\nu_{coadd} = \Psi_{coadd}/\sqrt{\Phi_{coadd}}$. In this $\nu$ image, generated for each given angle and velocity vector, any points above some threshold $m$ can be considered to be $m$-sigma detections of a moving source. For a single trajectory, we can define the summed likelihood as $\sum LH = \nu^{\mathrm{trajectory}}_{coadd}$. The interested reader is directed to \citet{kbmod} for more detail.

The large number ($\gg 10^9$) of potential asteroid trajectories means that these $\Psi_i$ and $\Phi_i$ images must be searched many times over. For this reason, KBMOD uses massively-parallel GPU computing for the core computations. The current software allows a user to search over $10^{10}$ potential moving object trajectories in a stack of 10-15 4K x 4K images in under a minute using a consumer-grade GPU (e.g., Nvidia 1080 Ti) \citep{kbmod}. Our pointer-arithmetic approach means that we never actually shift and stack images. Rather, we merely sum the previously-calculated likelihoods, utilizing thousands of concurrent GPU threads to keep the computation feasible on consumer-grade hardware.

The DECam NEO data set presents unique filtering challenges compared to \citet{kbmod} due to the increased number of potentially-valid trajectories, the short intra-night cadence, and image differencing artifacts. In \citet{kbmod}, detected sources appearing in the same position in 2 or more images, pixels with counts above 120 counts, and other mask flags set by the DECam community pipeline or the LSST software stack were all masked. In the current data set, we use difference imaging to subtract static sources. This enables us to decrease the masked area of the image, only masking sources flagged as detected if they appear in 10 or more images. However, despite reducing the number of detected individual sources on the image by a factor of about 18 (see Figure \ref{fig:sourcecomp}), leaving most of the image unmasked, coupled with difference imaging artifacts, increases the number of trajectories with $\sum LH > 10$ by a factor of 10. This problem is worsened by the intranight cadence. The average time between images within a single night is about 5 minutes. This means that for a characteristic trajectory with a velocity of 100 pixels per day, objects will move by less than 1 pixel between images. Conversely, this also means that if a static source appears along the potential trajectory, flux from this object will most likely be present in at least five trajectory data points, introducing repeated outliers into the trajectory.

\subsection{$\sigma_G$ Filtering}

In order to deal with the increased number of high likelihood trajectories (i.e. $10^7$ with $\sum LH > 10$), we developed faster, more-effective filtering. First, we altered how the GPU and C++ code handed off data to the Python-based filtering, leading to a speed increase of up to 300\%. Second, we replaced the Kalman filter used in \citet{kbmod} with a more statistically-robust quantile-based filtering method. We describe this new filtering method below.

With a traditional quantile-based filter, the filter rejects data points that are greater than $n\sigma$ from the central value of the distribution, where $\sigma$ is a measure of the spread of the distribution. In the case of a Gaussian distribution, $\sigma$ might be estimated by computing the standard deviation of the data and the central value estimated by computing the mean of the data. If we take $n=1$, then this simple filter would reject any data points that are greater than $1\sigma$ from the mean.

In the presence of significant outliers, the mean and standard deviation become biased estimators for the central value and the spread of the underlying Gaussian distribution. Following the approach of \citet{astroMLText}, we adopt a robust estimator for the central value and the true standard deviation of a Gaussian distribution with outliers. Consider the cumulative distribution function (CDF) of a Gaussian distribution

\begin{equation}
f(x) = \frac{1}{2} \left[ 1+ \mathrm{erf}\left(\frac{x-\mu}{\sigma_G \sqrt{2}}\right)\right]
\end{equation}

\noindent where $\mu$ is the mean, $\sigma_G$ is the standard deviation of the Gaussian, and $\mathrm{erf}$ is the error function. The inverse, then, is given by

\begin{equation}
x = \mu + \sigma_G \sqrt{2}\ \mathrm{erf}^{-1} \left[2 f(x) -1 \right]
\end{equation}

\noindent By sampling the Gaussian distribution at two quantiles $f(x_i)$ and $f(x_j)$, we can estimate $\sigma_G$. To do this, we take the difference of the inverted CDF

\begin{align}
    x_j - x_i &= \sigma_G \sqrt{2} \left( \mathrm{erf}^{-1} \left[2 f(x_j) -1 \right] -  \mathrm{erf}^{-1} \left[2 f(x_i) -1 \right] \right) \\
    \implies \sigma_G &= \frac{1}{ \mathrm{erf}^{-1} \left[2 f(x_j) -1 \right] -  \mathrm{erf}^{-1} \left[2 f(x_i) -1 \right]} \left(x_j - x_i\right) \\
    \implies \sigma_G &= C \left[ x_j - x_i \right]
\end{align}

\begin{figure}[tbh]
    \centering
    \includegraphics[width=1.0\textwidth,keepaspectratio]{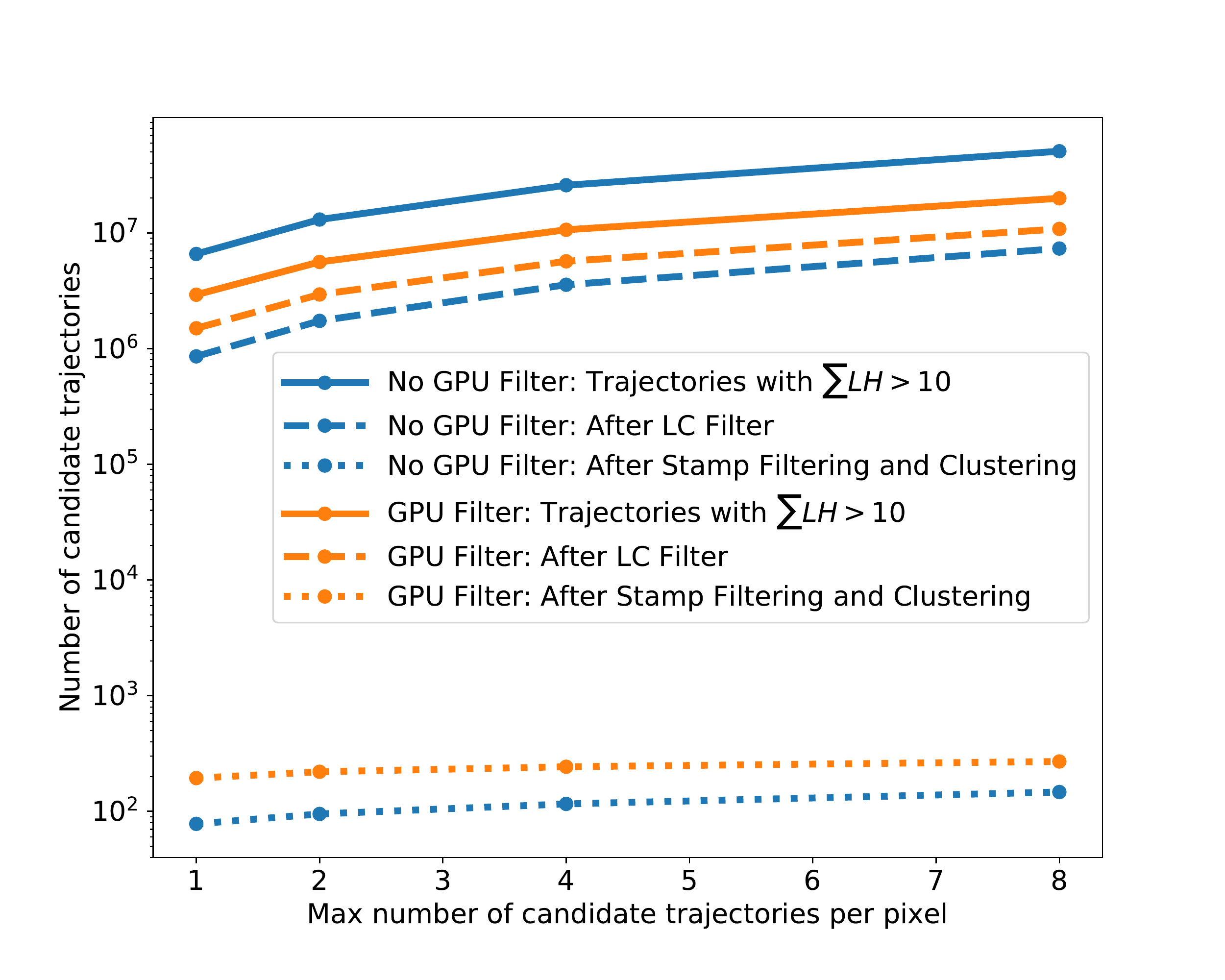}
    \caption{Number of candidate trajectories at various stages of processing for variable numbers of results per pixel. The solid line shows the number of candidate trajectories with $\sum LH > 10$ returned from the GPU for subsequent filtering. The dashed and dotted lines show the number of candidate trajectories passing CPU $\sigma_G$ filtering (dashed) and central moment stamp filtering and clustering (dotted). GPU filtering decreases the total number of candidate trajectories with $\sum{LH} > 10$, but increases the number of candidate trajectories that pass subsequent lightcurve filtering and stamp filtering and clustering. Because GPU memory constraints limit the number of candidate trajectories per starting pixel that can be saved for subsequent analysis, using a GPU filter means that the results that are passed out of the GPU are more likely to be potentially-valid. These results are then processed with the CNN filter and subject to human review. These data come from repeated reprocessings of pointing group 023, CCD 35.}
    \label{fig:res_per_pixel}
\end{figure}

\begin{figure}[tbh]
    \centering
    \includegraphics[width=0.9\textwidth,keepaspectratio]{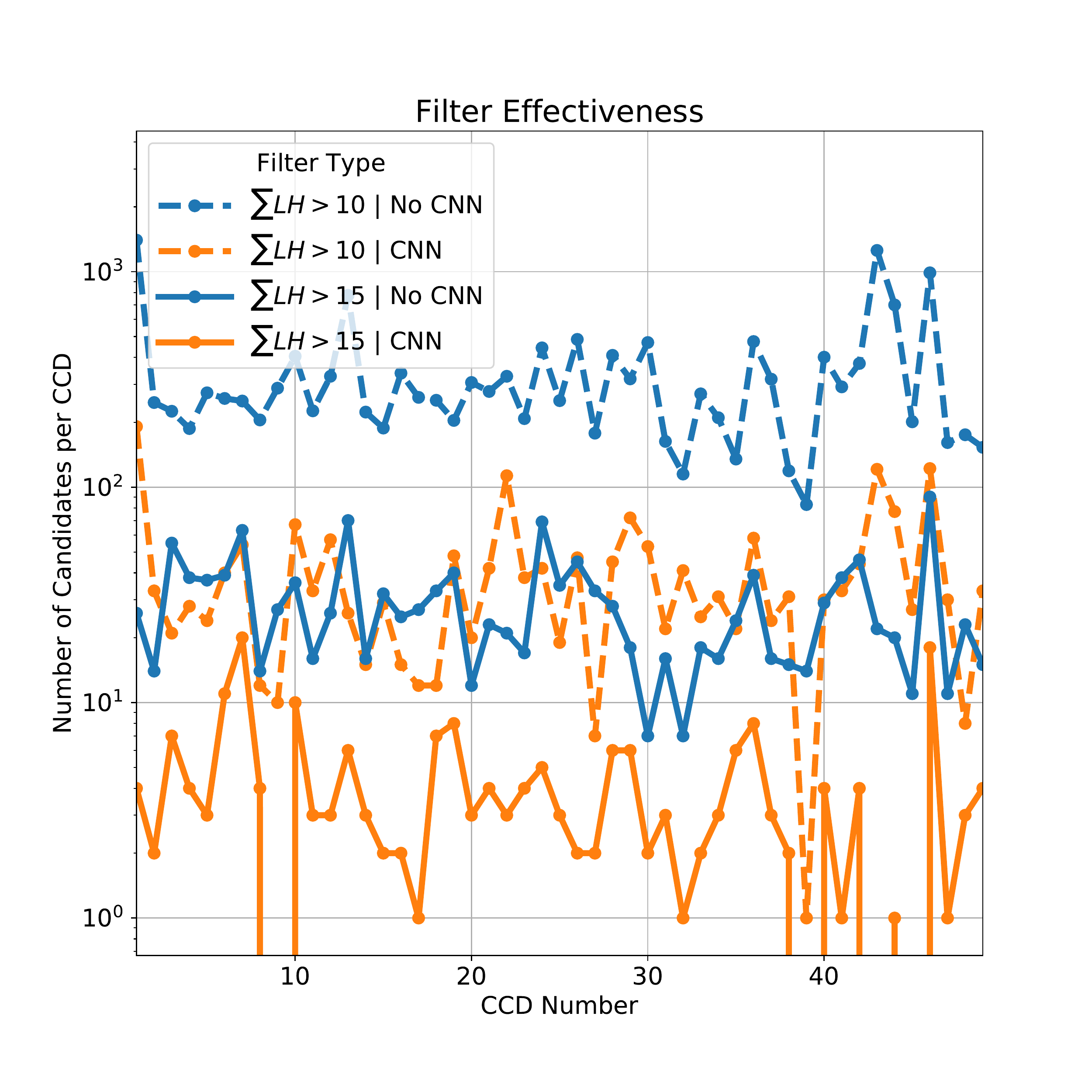}
    \caption{Number of results per CCD image stack from pointing group 190 requiring by-eye confirmation or rejection (hereafter ``candidate trajectories'') for likelihood limits of 10 (dashed line) and 15 (solid line), with (orange) and without (blue) RESNET 50 CNN filtering. These results are from pointing group 190, one of the search sample pointing groups. Here, the CNN was set to filter out any candidate trajectories with a probability of true that was less than 75\%. When using a LH limit of 15 and the CNN, the number of candidate trajectories per CCD was reduced to eleven or less, an acceptable number of trajectories for a human to review.}
    \label{fig:CNNStats}
\end{figure}

\begin{figure}[tbh]
    \centering
    \includegraphics[width=.97\textwidth,keepaspectratio]{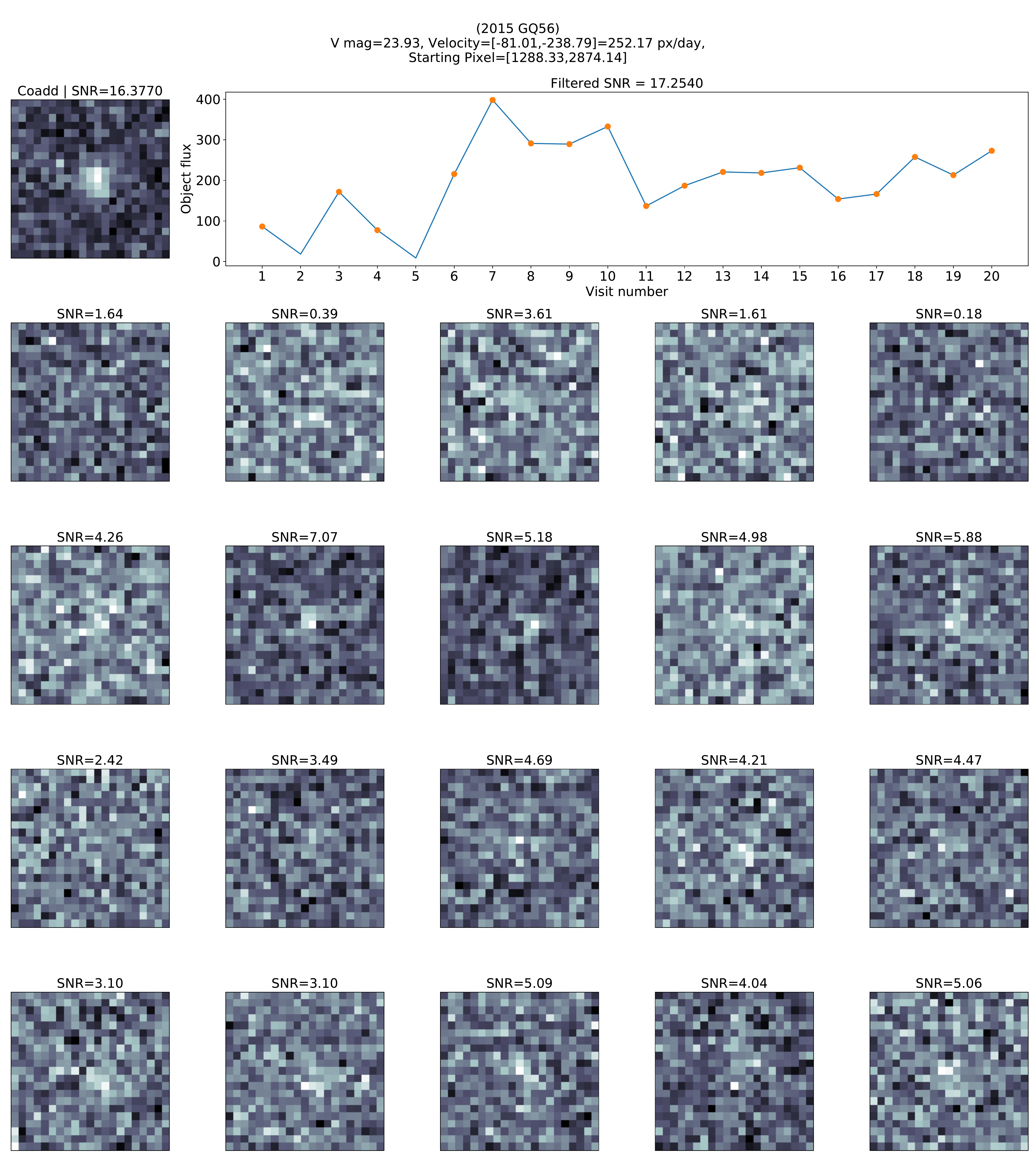}
    \caption{Sample output for object 2015 GQ56 (pointing group 300, CCD 30) when using trajectory estimates from the JPL Horizons service. The first row shows the coadded stamp (left) and the flux lightcurve (right). Orange points in the flux lightcurve are points that pass $\sigma_G$ lightcurve filtering. The remaining rows show the postage stamps for 2015 GQ56 in each individual image. The coadded stamp was generated by taking the median value at each pixel; this effectively removes image differencing artifacts. This trajectory was generated using orbital values from JPL Horizons. These figures were generated for all known KBOs in search sample in order to determine the unfiltered $\sum LH$, as well as for debugging purposes. Each stamp shows the estimated SNR $\nu$ of that stamp.}
    \label{fig:knownObject}
\end{figure}

Here, C is a coefficient dependent only on the choice of quantiles. $x_j$ and $x_i$ are estimated by selecting values from the lightcurve. The choice of upper and lower quantiles is user-determinable. Here, we estimate $\sigma_G$ using data from the 25th to 75th percentiles, for a coefficient of $C_{25,75} \approx 0.7413$. Then, we can estimate $x_{25}$ and $x_{75}$ from the data by selecting the 25th and 75th percentile values respectively from the data. We can then estimate the standard deviation of the underlying Gaussian distribution with $\sigma \approx 0.7413 \left( x_{75}-x_{25} \right)$. Given a robust estimator of the spread of the distribution (i.e. $\sigma_G$), we apply a filter that rejects any points that are not within $\pm n \sigma_G$ (e.g. $2 \sigma_G$) of the median of the data.

We apply this method to the likelihood and/or flux values of each trajectory. We then recompute $\sum LH$ for the trajectory values that pass the filter and reject the trajectory if the recomputed likelihood ($\sum LH'$) is less than 10. In practice, this filtering method successfully rejects of order $10^6$ erroneous candidate trajectories in approximately $60$s using 30 central processing unit (CPU) cores.

\subsection{In-line GPU Filtering}

Applying a variant of the $\sigma_G$ filter in the GPU while the search is running, instead of in post-processing, increases the number of potentially-valid trajectories returned to the CPU by KBMOD. In \citet{kbmod}, KBMOD passed the four trajectories per pixel with the highest $\sum LH$ from the GPU to the CPU. Other trajectories with the same starting pixel were discarded. Because KBMOD searches of order $10^{12}$ trajectories for a 2K x 4K image, it is computationally infeasible to keep the results of all evaluated trajectories in GPU RAM. The disadvantage of this approach is that lower-likelihood trajectories may get removed from the search even if they are valid trajectories of true objects. With reduced masking, there are many erroneous candidate trajectories with high likelihood. This means that removing the masks may have increased the probability of discarding valid trajectories.

In-line GPU filtering solves this problem by applying the filtering method to compute $\sum LH'$ before the trajectory is passed back to the CPU. This in-line GPU filter means that if a trajectory has a high $\sum LH$ only due to an outlier in the data, that trajectory is unlikely to supplant another valid trajectory when GPU results are passed back to the CPU. We also increased the number of returned results per pixel from four to eight. This means that we were able to process about four times as many results per pixel compared to \citet{kbmod}. The in-line GPU filtering uses a single GPU and is about 10\% faster than comparable CPU filtering using 30 CPU cores. Figure \ref{fig:res_per_pixel} demonstrates how the in-line GPU filter returns more potentially-valid trajectories for a given number of trajectories per pixel.

\subsection{Median Stamp Coadd Generation}

As shown in Figure \ref{fig:diffexp}, saturated cores and small image misalignments leave a number of artifacts in the difference image that also have to be accounted for in the filtering process. As in \citet{kbmod}, we computed the central moments of postage stamps for candidate trajectories. Stamps were rejected if they did not have central moments that were consistent with a Gaussian. In this data, we required that the x, y, xy, xx, and yy moments be strictly less than 0.5, 0.5, 1.5, 36.5, and 36.5 respectively. These values were chosen empirically based on the central moments of known KBOs. We generated coadded stamps by computing the median pixel value for each pixel along the trajectory. This mitigates the effect of image differencing artifacts, improving the performance of the central moment filter.

\begin{figure}[b!]
    \centering
    \includegraphics[width=1.0\textwidth,keepaspectratio]{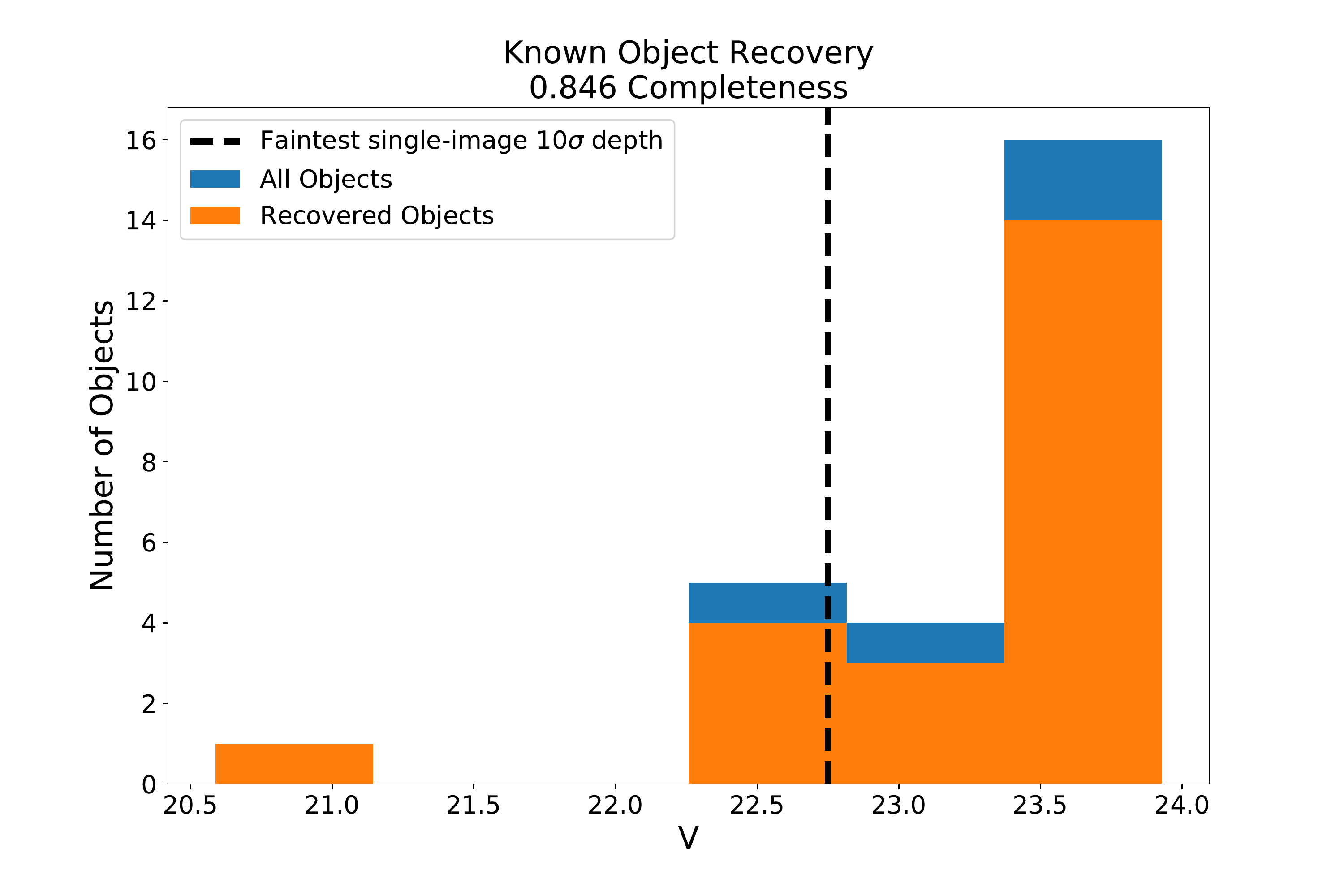}
    \caption{Recovered known objects as a function of reported magnitude. We ran an untargeted KBMOD search on all CCDs in the search sample that had known KBOs on them.  Figure \ref{fig:recoveryAnalysis} shows the recovery statistics for the recovered objects. 18 of the recovered objects were below the approximate upper-limit single-image $10\sigma$ limiting magnitude.}
    \label{fig:objectRecovery}
\end{figure}

\begin{figure}[tbh]
    \centering
    \includegraphics[width=1.0\textwidth,keepaspectratio]{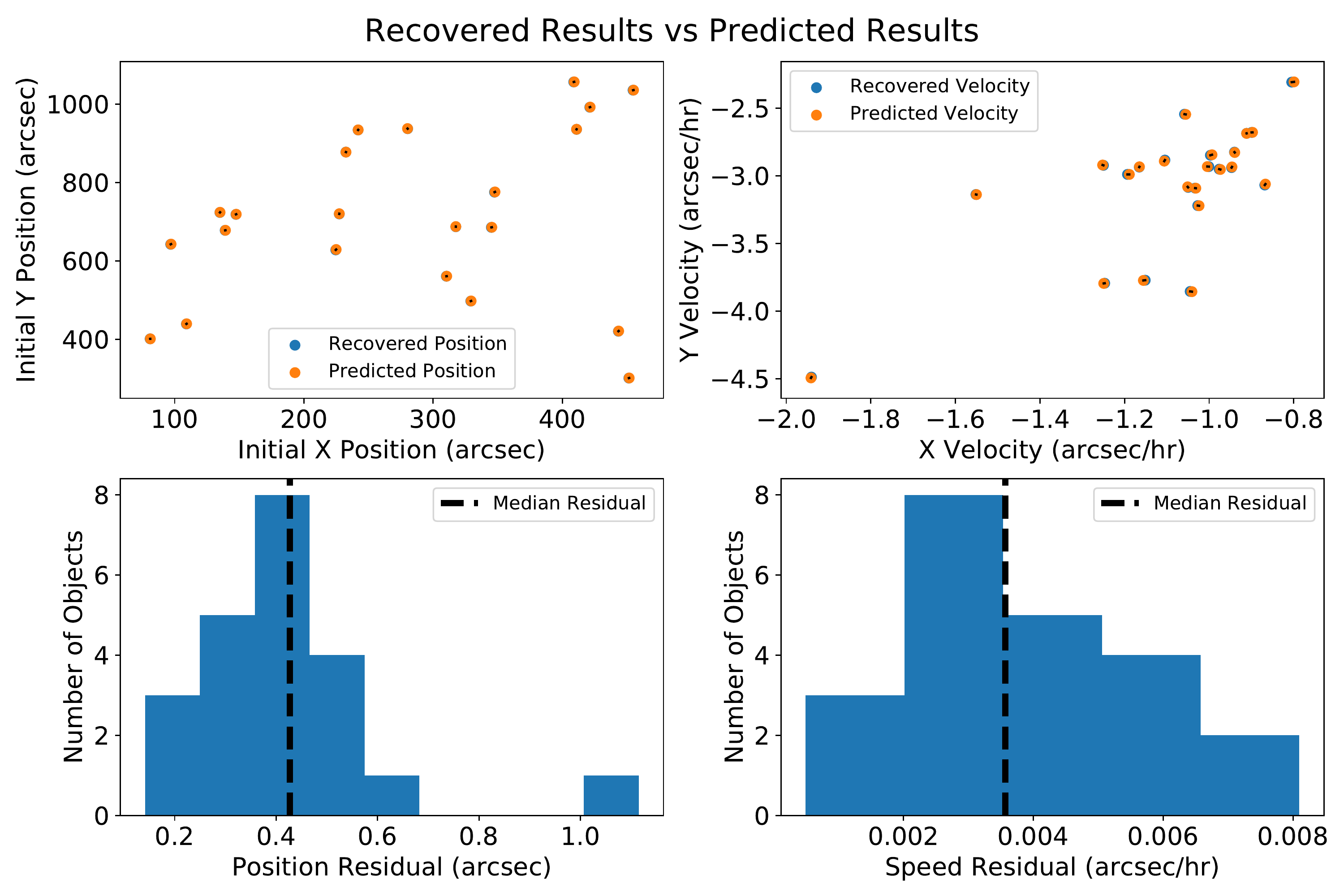}
    \caption{Statistics for the known objects that were recovered with a untargeted KBMOD search on CCDs with known objects in the search sample. For object recovery, we discarded any results that had a starting position more than 5 pixels (approximately $1.35"$, or one PSF FWHM) from the predicted location or had a velocity difference of more than 5 pixels per day (approximately $0.056"\ \mathrm{hr}^{-1}$). The velocity cutoff was chosen based on the recovery distribution. As shown in bottom left and bottom right respectively, the median difference between predicted and recovered position and speed was significantly lower than these cutoff values. The upper left plot shows each trajectory's initial predicted and recovered position on the CCD image for each object. The upper right plot shows each trajectory's predicted and recovered x and y velocity on the CCD image for each object.}
    \label{fig:recoveryAnalysis}
\end{figure}

\subsection{CNN Filtering}

To further reduce the number of false positives, we filter using a convolutional neural network (CNN). We built a Residual Network with 50 layers (ResNet50\footnote{\url{https://github.com/priya-dwivedi/Deep-Learning/blob/master/resnet_keras/Residual_Networks_yourself.ipynb}}) \citep{he2015deep}. Residual networks are a type of CNN that add ``shortcut connections'' into the network architecture, which help to train deeper networks. Training a CNN requires a large amount of representative data. In this case, we needed a large ($>10^4$) labeled set of 21x21 stamps containing approximately equal numbers of false positives and true positives. To generate false positives, we ran an untargeted search (with similar grid spacing as described in \ref{sec:search-detect-recovery}) with a coadded likelihood limit of $\sum LH > 10$ along trajectories unlikely to correspond to real objects (approximately $90^\circ$ from the direction of the ecliptic). We ran a total of 53 searches with data from 34 unique pointing groups. These pointing groups were not constrained to the search sample. These searches yielded 113,549 21x21 false positive postage stamps. Because KBOs are relatively rare, we could not use real recovered objects to generate the thousands of true positives needed to train the CNN. To circumvent this limitation, we generated 44,950 simulated true positives. To make these stamps, we retrieved 25 21x21 postage stamps from a CCD along a semi-random trajectory. Next, we drew a random brightness from an exponential distribution (with dimmer objects being the most likely). Using this brightness, we added a Gaussian to each background stamp with a random standard deviation ($1-2.1$ pixels), a random central offset ($<2$ pixels), and a random linear offset ($<2$ pixels over the image time baseline). To train the CNN, we cut the false positive stamps and simulated true stamps down to 40,000 randomly-selected coadded stamps each. We used 70\% of the data for training, 20\% for validation, and the remaining 10\% for testing. After 20 epochs, the training set accuracy was about 99\%, while the validation set accuracy was about 96\%. After training, the test set accuracy was also about 96\%.

This CNN returns a predicted probability that a coadded postage stamp contains a simulated object. Because the stamps of simulated objects differ from the stamps of real objects, this probability is not a perfect representation of the likelihood that a coadded stamp contains a real object. However, it creates a user-programmable threshold that can be used to reduce false positives enough that the remaining candidate trajectories can be analyzed by-eye. We reject any stamps with a CNN probability of true less than 75\%. As shown in Figure \ref{fig:CNNStats}, when reviewing only objects with a $\sum LH>15$ and using this CNN filter, there are generally fewer than 10 candidate trajectories per CCD that require human by-eye confirmation or rejection.

\section{Results} \label{sec:results}
\subsection{Search, Detection, and Recovery} \label{sec:search-detect-recovery}
\begin{figure}[tbh]
    \centering
    \includegraphics[width=0.93\textwidth,keepaspectratio]{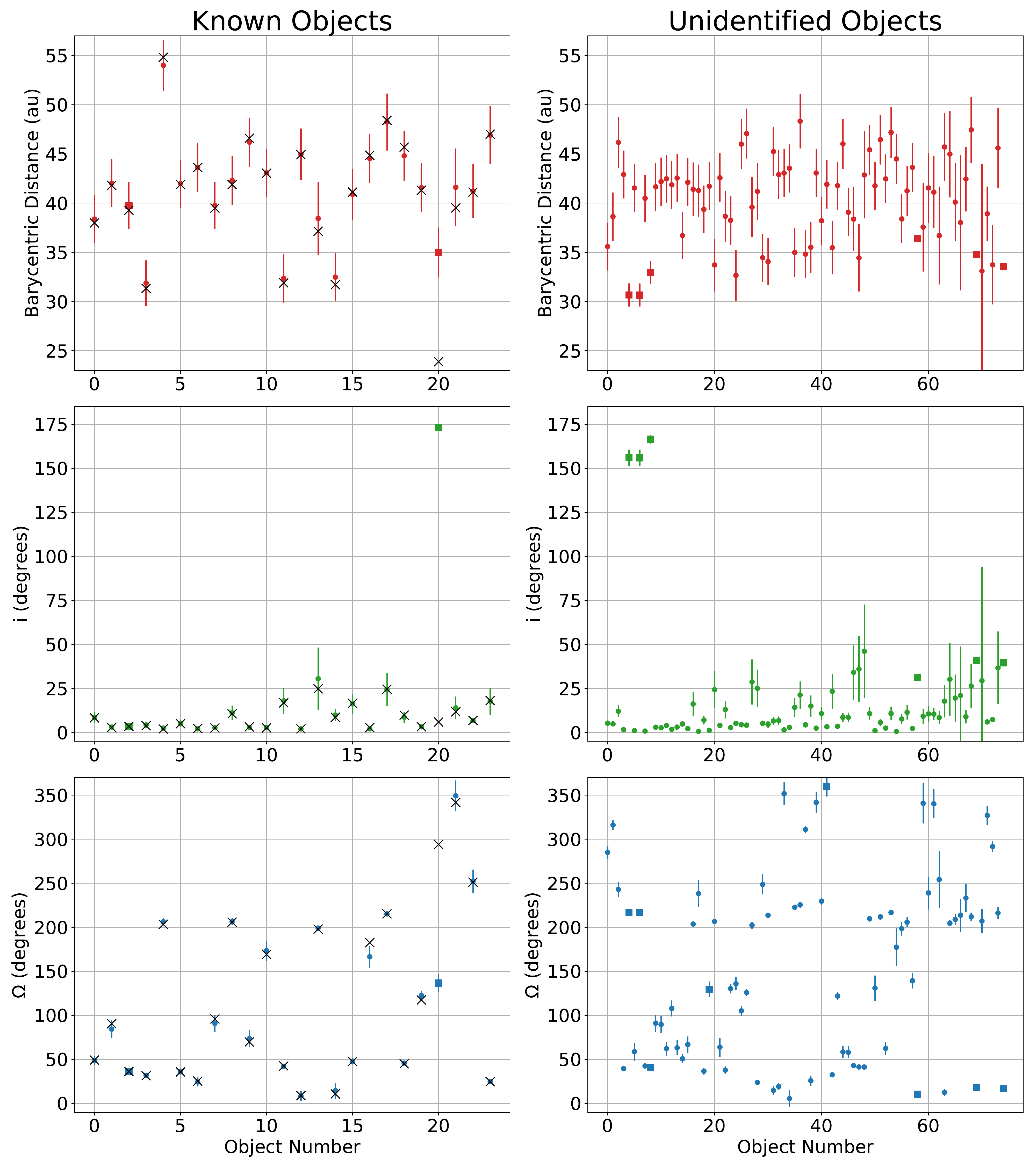}
    \caption{Best-fit barycentric distance $r_0$, inclination $i$, and longitude of ascending node $\Omega$ (dots) with respective standard deviations (lines) of the detected known objects (left) and unidentified objects (right) using the method of \citet{Bernstein2000}. $r_0$, $i$ and $\Omega$ were also fit with Find\_Orb. When the value from Find\_Orb is inconsistent with \citet{Bernstein2000} within 1$\sigma$, we show the best-fit value from \citet{Bernstein2000} with a square instead of a dot. For the known objects, the JPL Horizons value of the corresponding parameter is overplotted with an x marker. The short time baseline of the observations allows us only to constrain initial barycentric distance, inclination, and longitude of ascending node. The medians of the absolute value of the residuals between the best-fit values and the JPL Horizons values are 0.36 au, 0.32 degrees, and 0.92 degrees for $r_0$, $i$, and $\Omega$ respectively. As reported by JPL Horizons, the median values of the known objects for $r_0$ and $i$ are $\widetilde{r_0} = 41.55$ au and $\widetilde{i} = 5.46^\circ$ respectively. The median values of the unidentified objects for the best-fit $r_0$ and $i$ are $\widetilde{r_0} = 41.28$ au and $\widetilde{i}=7.67^\circ$. }
    \label{fig:orbfit}
\end{figure}

We ran an untargeted KBMOD search on each stack of CCDs in the search sample for a total of 2580 searches. Similar to \citet{kbmod}, an untargeted search looks for linear trajectories with velocities between 92 and 550 pixels per day ($1.04"\ \mathrm{hr}^{-1}$ to $6.19"\ \mathrm{hr}^{-1}$) with angles of $\pm \pi/10$ from the ecliptic angle. Compared to \citet{kbmod}, we doubled the resolution of the grid spacing from 256 velocity steps and 128 angle steps to 512 velocity steps and 256 angle steps. This ensured that trajectories would end up separated by no more than about two PSF FWHM from neighboring trajectories.

In order to test the efficiency of these new filtering methods, we generated a list of known objects in the search sample. We used Skybot \citep{Skybot} and JPL Horizons \citep{JPL_Horizons} to find all KBOs that were present in the search sample, with the additional requirement that they be present in the first image of the pointing group. We generated 21x21 pixel postage stamps of the object in each image in which it is present. We developed a variant of KBMOD that computes the likelihoods along a single trajectory then runs the aforementioned quantile-based filtering, and computed the central moments of the postage stamps.  Figure \ref{fig:knownObject} shows these results for pointing group 300, CCD 30, object 2015 GQ56. We removed KBOs with an unfiltered $\sum LH < 15$. This left us with a ``recovery sample'' of 26 KBOs.

In the untargeted search of the search sample, we recovered 22 out of 26 (or 84.6\%) of the known objects in the recovery sample after all filtering was applied (see Figure \ref{fig:objectRecovery}). The CNN probability threshold was kept at 75\%. Recovery statistics for these objects are shown in Figure \ref{fig:recoveryAnalysis}. For object recovery, we discarded any trajectories that had a starting position more than 5 pixels (approximately $1.35"$, or one PSF FWHM) from the predicted location or had a velocity difference from the known velocity of more than 5 pixels per day (approximately $0.056"\ \mathrm{hr}^{-1}$). The median position and speed residuals were $0.427"$ and $0.0036"\ \mathrm{hr}^{-1}$ respectively, significantly below the chosen cutoff values. This velocity error corresponds to approximately a 1.27 pixel position error over four days. Using the NOAO DECam Exposure Time Calculator (ETC), we estimate the single-image $10\sigma$ depth to be at most $22.75$V. Because the pointing groups contain data from different nights, we computed this limit assuming a new Moon. It is therefore an upper limit. 18 of the recovered objects were fainter than the upper-limit single-image $10\sigma$ depth. This confirms that KBMOD is able to use difference images to find moving KBOs that are too dim to detect in a single image at the $10\sigma$ level, extending the result of \citet{kbmod} to difference images.

We investigated each of the missed known objects individually. 2013 GY136 (pointing group 204, CCD 57) failed to process due to a CCD that failed image differencing. This reduced the total number of images in CCD 57 to fewer than 20, and CCD 57 was therefore not reprocessed. 2013 GZ137 (pointing group 202, CCD 52) failed CNN filtering with a threshold of 75\%, but passes with a threshold of 50\%. 2015 GY55 (pointing group 306, CCD 26) starts within 4 pixels of the chip edge, causing this trajectory not to be searched by KBMOD. 2013 GH137 (pointing group 192, CCD 41) has two fully-masked stamps, and two more with partial masking, which may have caused it to be filtered out.

In addition to the detected 22 known objects in the recovery sample, we detected 2 additional known KBOs. These KBOs had an unfiltered $\sum LH < 15$ along the JPL Horizons trajectories, and were therefore not included in the recovery sample. The best KBMOD trajectories for these objects had a filtered $\sum LH' > 15$. We then linked these objects back with known KBOs.

\begin{figure}[tbh]
    \centering
    \includegraphics[width=1\textwidth,keepaspectratio]{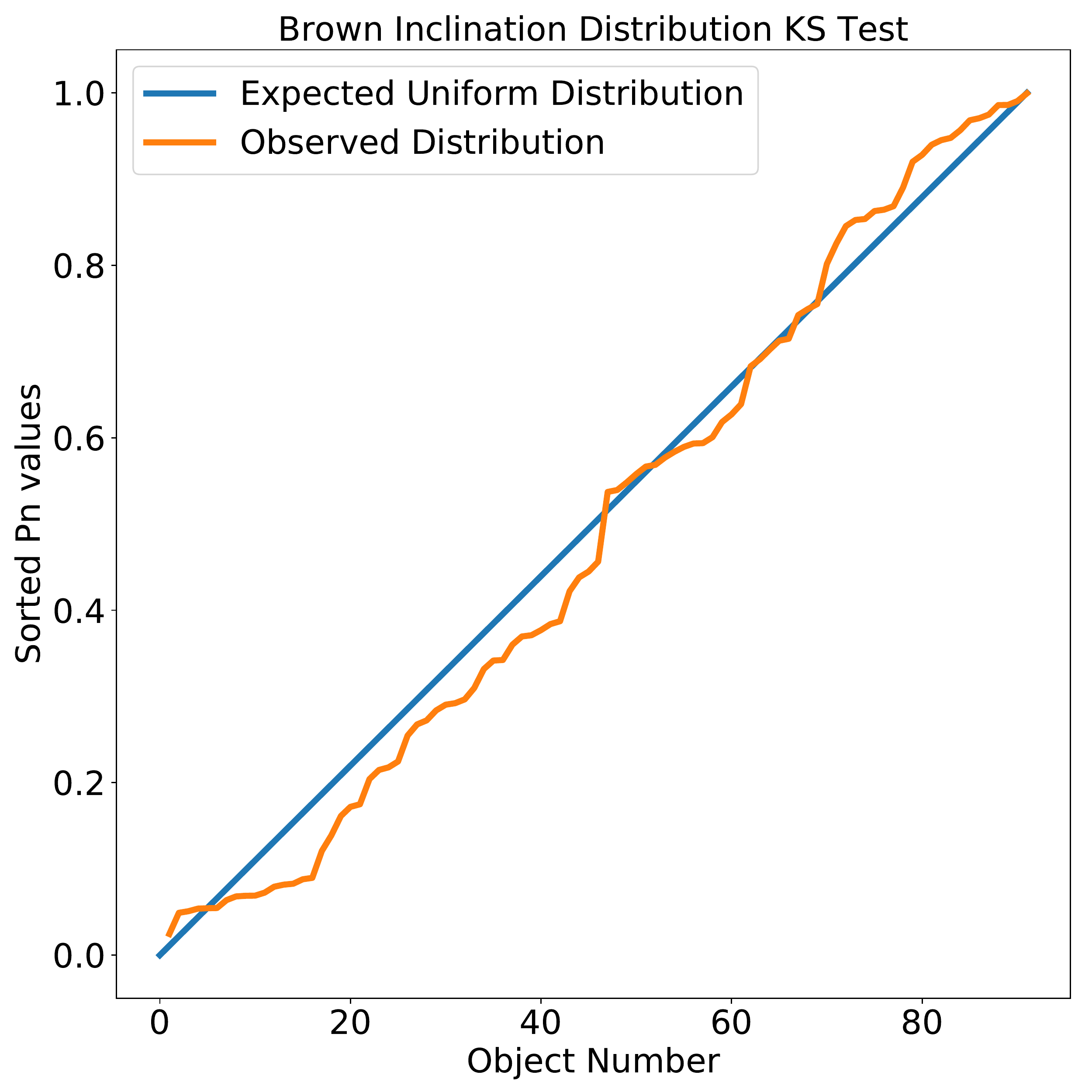}
    \caption{One-sided Kuiper variant of the Kolmogorov-Smirnov (K-S) test comparing our recovered inclinations with the inclination distribution predicted by \citet{brown_inclinations}. We reject the null hypothesis that our inclinations came from the distribution of \citet{brown_inclinations} with only 76.6\% confidence (less than 1$\sigma$). We therefore consider our observed inclinations to be consistent with the distribution predicted by \citet{brown_inclinations}.}
    \label{fig:KS_inc}
\end{figure}

\subsection{Orbit Fitting and Analysis}
We detected 75 moving objects that we were unable to link to existing objects. Trajectories with $\sum LH > 15$ that passed all filtering were accepted or rejected with a by-eye examination of the individual stamps, the coadded stamp, and the flux lightcurve.

As shown in Figure \ref{fig:orbfit}, we used the method described in \citet{Bernstein2000} to fit barycentric distance $r_0$, inclination $i$, and longitude of ascending node $\Omega$ of both the recovered known objects and the unidentified objects. For the known objects, we compared the orbital parameters fit to the KBMOD trajectory with their respective parameters as reported by JPL Horizons. The medians of the absolute value of the residuals between the best-fit values and the JPL Horizons values are 0.36 au, 0.32 degrees, and 0.92 degrees for $r_0$, $i$, and $\Omega$ respectively. The median values for $r_0$ and $i$ of the known objects reported by JPL Horizons are $\widetilde{r_0} = 41.55$ au and $\widetilde{i} = 5.46^\circ$ respectively. The median values of the unidentified objects for the best-fit $r_0$ and $i$ are $\widetilde{r_0} = 41.28$ au and $\widetilde{i}=7.67^\circ$. The three parameters (shown in Figure \ref{fig:orbfit}) that are well-fit with our data constrain the plane of the orbit and the initial distance of the object from the Solar System barycenter. Individual values are shown in Table \ref{sec:param_table}.

In addition to the method of \citet{Bernstein2000}, we used Find\_Orb\footnote{\url{https://github.com/Bill-Gray/find_orb}} to fit $r_0$, $i$, and $\Omega$. This allowed us to compare the best-fit values between the two orbit fitting codes. When best-fit values from Find\_Orb were not within 1$\sigma$ of the best-fit value from the \citet{Bernstein2000} code, we show the value as a square in Figure \ref{fig:orbfit}. We discarded values with inconsistent inclinations from the remainder of the orbit analysis.

There were a few noteworthy limitations to our dataset and apparent outliers in our best-fit values. Because of the relatively short time baseline of about four days, we were unable to place any meaningful constraints on the other Keplerian elements individually. For three unidentified objects (unidentified object numbers 58, 69, and 74), the orbit fitting code did not return uncertainties. We therefore consider them inconsistent between \citet{Bernstein2000} and Find\_Orb. Unidentified object numbers 4, 6, and 8 have a best-fit inclination of $i_{\mathrm{fit}}> 90^\circ$. Similarly, known object number 20 (2000 EE173) has a best-fit inclination of $i_{\mathrm{fit}} = 173.36 ^ \circ \pm 0.54 ^ \circ$, but a JPL Horizons inclination of $i_{\mathrm{Horizons}}=5.95^ \circ$. However, these 4 objects are all marked as inconsistent between Find\_Orb and \citet{Bernstein2000}. As such, their best-fit values are removed from further orbital analysis.

\begin{figure}[tbh]
    \centering
    \includegraphics[width=1\textwidth,keepaspectratio]{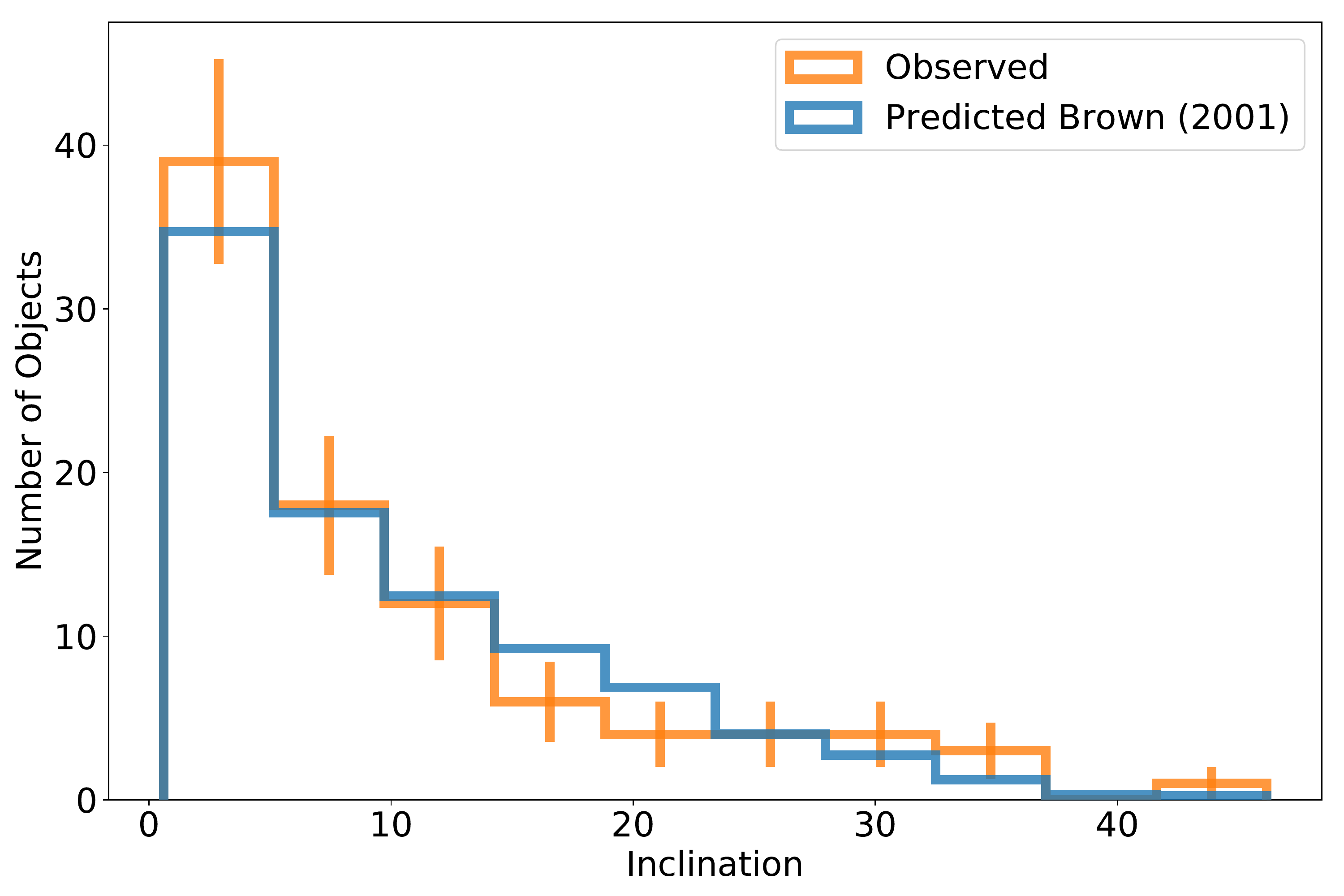}
    \caption{Inclination distributions of our detected objects (orange) and the distribution predicted based on \citet{brown_inclinations} (blue), after accounting for the search sample ecliptic latitudes. The \citet{brown_inclinations} is weighted to the number of objects recovered by KBMOD. Uncertainties in the orange histogram are calculated as 1$\sigma$ Poisson intervals of $\sqrt{N}$.}
    \label{fig:expected_inc}
\end{figure}

To evaluate the consistency of the properties of our detected asteroids with published distributions, we apply the analysis of \citet{kbmod} to the detected objects with consistent inclinations reported in this paper. We compared our observed inclination distribution with that of \citet{brown_inclinations} by using a one-sided Kuiper variant of the Kolmogorov-Smirnov (K-S) test. We use a test statistic of $D\sqrt{N}$ where $N$ is the number of objects, and $D$ is given by Equation 30 in \citet{kbmod}.

\begin{equation}
    D=\mathrm{max}\left(P_j-j/N\right)
\end{equation}

\noindent$P_j$ is the probability for a given inclination distribution that an object $j$ has an inclination equal to or below the actual inclination $i_j$. Some TNO sub-populations have non-uniform inclination distributions around the ecliptic. This is an unmodeled systematic in our test statistic. We compute $P_j$ using Monte Carlo methods. We take $10^5$ inclinations from the \citet{brown_inclinations} distribution, place them randomly along circular orbits and take all objects within $\pm0.5^\circ$ of the ecliptic latitude $\beta_j$ of discovery. These values allow us to find $P_j$ by calculating the probability that an object with a given $\beta_j$ has an inclination at or below $i_j$. We run 1000 Monte Carlo simulations, using the mean $D\sqrt{N}$ as our test statistic. See Section 4.2.1 of \citet{kbmod} and Section 3 of \citet{brown_inclinations} for more detail.

Our mean value for $D\sqrt{N}$ was 1.40. As shown in Figure \ref{fig:KS_inc}, we reject the null hypothesis that our observed inclinations come from the distribution of \citet{brown_inclinations} with only 76.6\% confidence, which is less than the 1$\sigma$ confidence level of 84.1\% ($D\sqrt{N} = 1.47$). This is to say that we cannot confidently reject the null hypothesis. We can therefore say that our observed inclinations are consistent with \citet{brown_inclinations}.

We repeated the further comparison of \citet{kbmod}, using an approximate survey simulation to identify the distribution of objects with a given inclination that we would expect to find given the central RA and Dec of our search sample. We modeled the DECam field of view as a circle with a diameter of $2.2^\circ$. We used the inclinations and orbits from the Monte Carlo simulations used to generate Figure \ref{fig:KS_inc} and recorded the objects visible within the simulated camera footprint. We then normalized this simulated object distribution to the number of detected objects in the search sample. Figure \ref{fig:expected_inc} shows the simulated distribution (blue) and the observed distribution (orange). The $\chi^2$ value between the simulated and expected distributions was 8.58, corresponding to a $p$-value of 0.48. We therefore again say that our observed inclinations are consistent with \citet{brown_inclinations}.

\begin{figure}[tbh]
    \centering
    \includegraphics[width=1\textwidth,keepaspectratio]{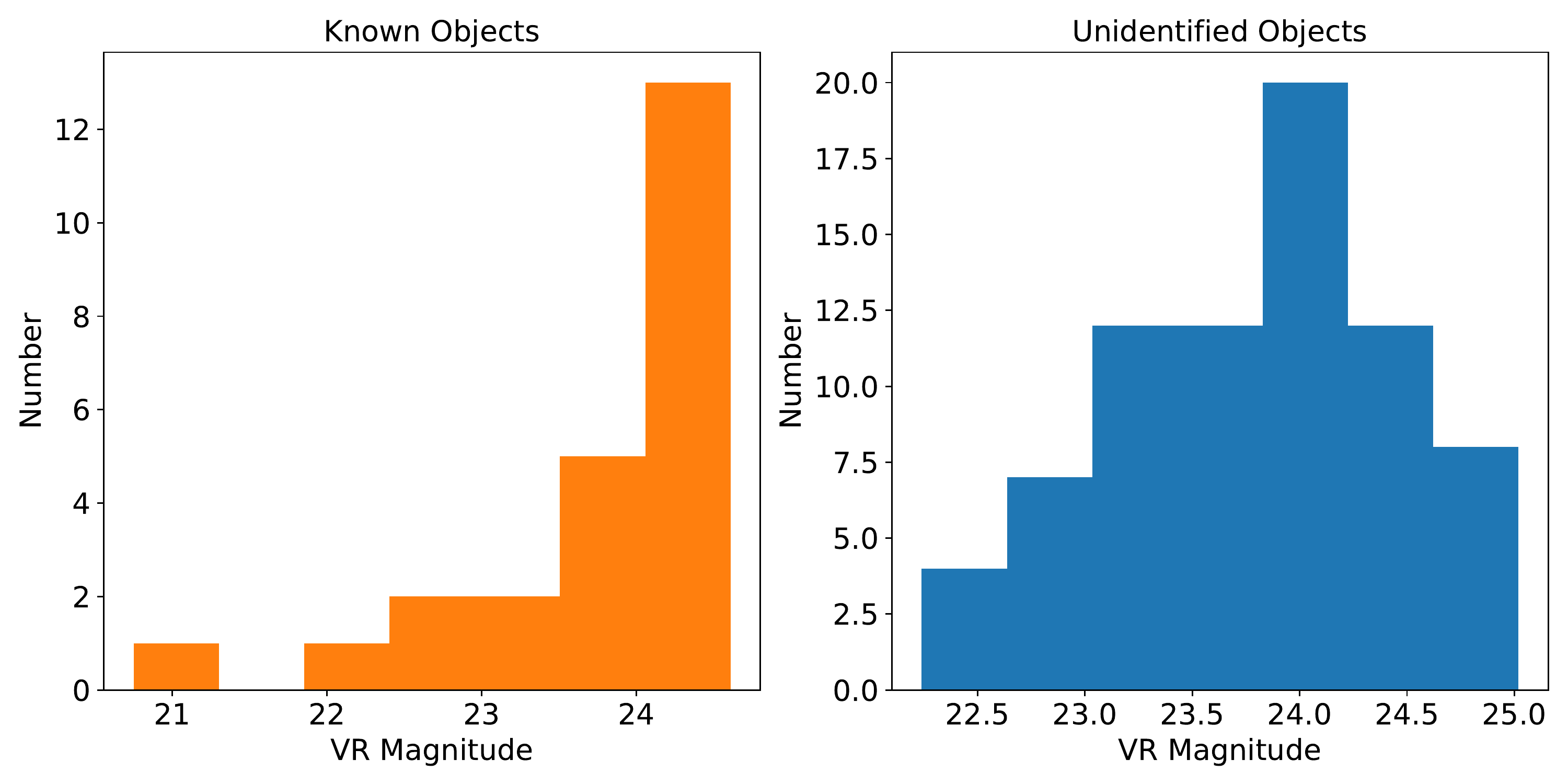}
    \caption{Best-fit \textit{VR} magnitudes for the previously-known objects (left) and unidentified objects (right).}
    \label{fig:VRmag}
\end{figure}

\begin{figure}[tbh]
    \centering
    \includegraphics[width=1\textwidth,keepaspectratio]{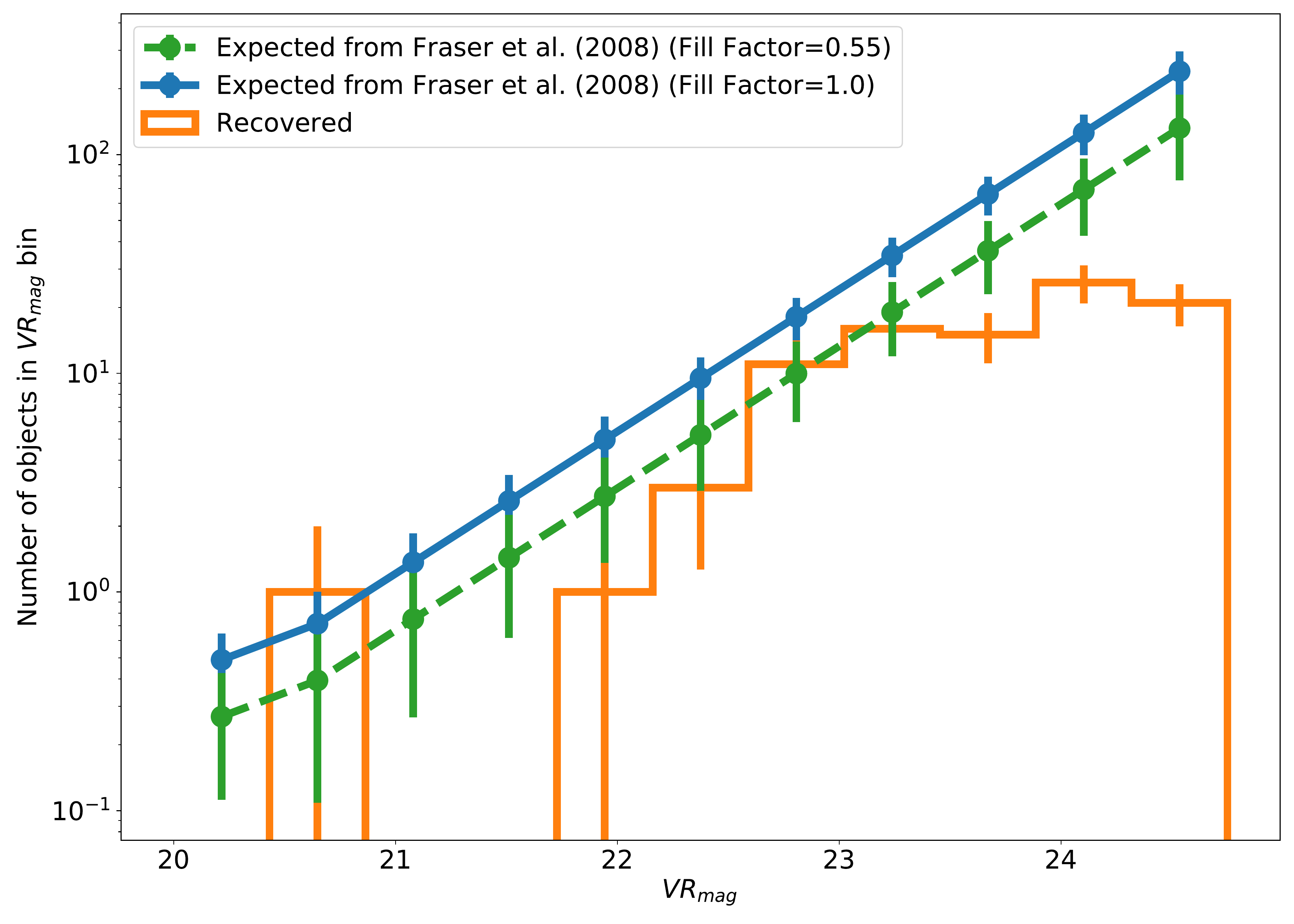}
    \caption{\textit{VR} magnitude distribution (orange) of recovered objects (known and unidentified), along with the number of objects predicted by \citet{FRASER2008827} assuming a circular camera footprint of 3 square degrees with no fill factor (blue) and a fill factor of 0.55 (green). Uncertainties in the orange histogram are calculated as 1$\sigma$ Poisson intervals of $\sqrt{N}$.}
    \label{fig:mag_compare}
\end{figure}

\subsection{Magnitude Estimation and Analysis}\label{sec:mag}
Figure \ref{fig:VRmag} shows our estimates of the \textit{VR} magnitude of the known and unidentified objects detected with KBMOD. To fit the \textit{VR} magnitudes, we generated 25x25 pixel postage stamps in the undifferenced science images following the KBMOD linear trajectory. In each stamp, we fit for the location of the object by maximizing the value of the flux minus the stamp background. The flux was calculated by summing the counts within a circular top-hat psf with a radius of twice the FWHM of the stamp. The local stamp background was estimated from the region outside of this psf. The magnitude zero point was obtained from the InstCal images. We then took the median magnitude value from each set of 15 to 20 magnitude estimates.

As we did in \citet{kbmod}, we compared our joint magnitude distribution with the apparent magnitude luminosity function presented in \citet{FRASER2008827}, adjusting for ecliptic latitude by using the inclination distribution of \citet{brown_inclinations}. We use the $<VR-R>$ KBO color reported by \citet{FRASER2008827}. We note, however, that the DECam \textit{VR} filter of our observations differs somewhat from the Mosaic2 \textit{VR} filter used in \citet{FRASER2008827}. They have similar central wavelengths, but different filter response curves. Individual magnitudes are shown in Table \ref{sec:param_table}. The magnitude uncertainties listed in Table \ref{sec:param_table} are reported as $\sigma_G$ uncertainties estimated from each set of magnitude estimates.

We approximate the camera footprint as a 3 square degree circle. In practice, our trajectories do not cover the entire camera footprint. Each individual KBMOD search only uses data a single CCD, requiring 60 individual searches to cover a full camera footprint. We further require that each candidate trajectory have at least 15 observations, corresponding to a time baseline of about 3 days. Depending on the search velocity and angle, this means that any objects that start near a CCD edge will not be searched, as the trajectory will go off the CCD edge before the trajectory has the requisite 15 observations. We define an effective search fill factor as the fraction of the CCD that is actually searched with KBMOD. For our search parameters, the search fill factor varies from about 0.5 to about 0.9. Assuming a typical KBO speed and angle of 275 pixels per day with an in-image angle of 4.4 radians gives a typical search fill factor of around 0.7. Multiplying this by the camera active-pixel fill factor of about 0.8 gives a typical net fill factor of approximately 0.55.

Figure \ref{fig:mag_compare} shows a histogram of our observed \textit{VR} magnitudes along with the number expected from \citet{FRASER2008827} assuming a fill factor of 1.0 and 0.55. Our joint magnitude distribution is largely inconsistent with \citet{FRASER2008827} assuming a fill factor of 1.0, but is consistent to within uncertainties up to about $VR=23.25$ assuming a fill factor of 0.55.

\section{Discussion} \label{sec:discuss}

The improvements already presented in this paper helped enable KBMOD to detect 22 out of 26 known objects in the recovery sample. The trajectories of these known objects were recovered with a median error in starting position of less than two pixels. Furthermore, KBMOD was able to detect 75 objects that we were unable to link with any previously-known objects. Although the time baseline of the data was short, we were able to fit the barycentric distance, inclination, and longitude of ascending node of both the known objects and the unidentified objects. The inclination distribution of the recovered objects is consistent with the distribution from \citet{brown_inclinations}. The number of objects detected as a function of magnitude is consistent with the distribution from \citet{FRASER2008827} assuming a net fill factor of 0.55.

\citet{kbmod} validated KBMOD on the High Cadence Transient Survey (HiTS) \citep{HITS}. This work validates algorithmic improvements to KBMOD filtering with a survey that has a time baseline of up to four nights, compared to the three nights used in \citet{kbmod}. Furthermore, this work validates KBMOD as applied to images that have been differenced with a coadded template.

In so doing, we demonstrated that KBMOD can recover KBOs in difference images from a survey with a longer time baseline and an irregular cadence. However, this required more robust filtering methods. By adding GPU filtering, we have increased the effective number of potentially-valid candidate trajectories that can be passed out of the GPU for further filtering and analysis. With the $\sigma_G$-based filtering, we have also implemented more robust lightcurve filtering that improves filtering with an irregular image cadence. The CNN ResNet50 stamp filter shows great promise for future stamp filtering methods.

Next-generation astronomy surveys will soon be current-generation. This imminent wealth of data will require new computational tools in order to access its full potential. KBMOD has the potential to increase the number of TNOs detected with LSST from $\sim 40,000$ to $\sim 320,000$ as well as investigate the faint and mysterious class of objects at the very edge of our Solar System. In terms of probing the sednoids, with three months of coadded data we could detect a Sedna-like object at opposition at over 290 au, as opposed to $\sim 210$ au for a single image. With a year of coadded data, 290 au increases to 310 au. If we could coadd the entire LSST survey, 310 au increases to over 415 au. Note that objects on elliptical orbits spend much more of their time further from the Sun. If Sedna, which was detected near its perihelion around 90 au \citep{Brown_2004}, is representative of a larger population of sedoids, then most of these objects should be closer to apocenter than pericenter. Therefore, a linear increase in detection distance should yield a super-linear increase in the number of detected objects on a similar orbit. With this coaddition approach, it might even be possible to detect inner Oort Cloud objects with perihelion near 400 au, and aphelion well beyond.

Further work is needed before KBMOD will be able to run on LSST. We do not currently address the ``look-elsewhere'' effect in our search algorithm \citep[e.g.][]{look_elsewhere}. However, our false positives are already dominated by image artifacts and real sources. Even after filtering, trajectories require human by-eye confirmation or rejection. Because of this requirement of human review, we consider this an acceptable limitation. Future work will further investigate necessary algorithmic improvements to enable machine-only object confirmation, including addressing the ``look-elsewhere'' effect.

Enabling KBMOD to search across multiple CCDs will increase the effective fill factor, enabling greater completeness and longer time baselines. CCD chip gaps and camera edges will always keep the fill factor below 1.0 (relative to a circular footprint). However, with a CCD chip gap between 153 (columns) and 201 (rows) pixels, a KBO would move past the chip gap and onto the next CCD in about one night, assuming a typical KBO velocity of 275 pixels per day.

Improving image astrometry and image differencing is likely to reduce the number of image differencing artifacts, thereby reducing the number of candidate trajectories requiring by-eye detection. The non-uniformity of the image time baseline in this survey means that artifacts appeared in approximately the same location in up to five images. This posed a unique challenge to filtering out artifacts from candidate trajectories. Because of these factors, and because we ultimately validate each detected object by-eye, we save a full efficiency analysis of KBMOD for a future survey.

Given the relatively low inclination (median value of $\widetilde{i} = 7.67^\circ$) and barycentric distances between 30 au and 50 (median value of $\widetilde{r_0} = 41.28$ au), we find it likely that the majority of the unidentified objects presented in Figure \ref{fig:orbfit} are Kuiper belt objects. However, because the short arcs prevent us from placing accurate constraints on semi-major axis and eccentricity, we are unable to confirm this prediction with the current data. Future follow-up or precovery attempts for these objects may be able to extend the observational arcs enough to accurately constrain them to the Kuiper belt, and perhaps place them within a Kuiper belt subpopulation (e.g. the cold classical Kuiper belt).

The authors acknowledge support from NASA awards NNG16PJ23C and 80NSSC21K1528, and NSF awards AST-1715122, AST-1409547, and OAC-1739419. This work used the Extreme Science and Engineering Discovery Environment \citep[XSEDE; ][]{XSEDE}, which is supported by National Science Foundation grant number ACI-1548562. This work used the XSEDE Bridges GPU and Bridges-2 GPU-AI at the  Pittsburgh Supercomputing Center through allocation TG-AST200009. The authors acknowledge support from the DIRAC Institute in the Department of Astronomy at the University of Washington. The DIRAC Institute is supported through generous gifts from the Charles and Lisa Simonyi Fund for Arts and Sciences, and the Washington Research Foundation.

\software{KBMOD \citep{kbmod}, LSST Science Pipelines \citep{LSST_DM}, astropy \citep{astropy}, scikit-image \citep{skimage}, numpy \citep{numpy}, CUDA \citep{cuda}, scikit-learn \citep{scikit}, pandas \citep{pandas}, matplotlib \citep{matplotlib}, tensorflow \citep{tensorflow}}
\bibliography{references}
\appendix

\section{Table of Detected Object Parameters}\label{sec:param_table}

\begin{longtable}{llllll}

\toprule
\Longunderstack{Identifier\\(pg, ccd)} &    \textit{VR} Mag & \Longunderstack{Barycentric\\Distance\\(au)} &      i (degrees) & $\Omega$ (degrees) &  \Longunderstack{Linked to\\Known Object} \\

\endhead

\multicolumn{6}{r}{{Continued on next page}} \\

\endfoot

\endlastfoot
            (190,20) &  24.46$\pm$0.22 &            38.38$\pm$2.42 &    8.58$\pm$2.99 &     48.73$\pm$4.99 &                    True \\
            (190,23) &  24.13$\pm$0.82 &            42.02$\pm$2.43 &    3.10$\pm$0.50 &    84.35$\pm$10.33 &                    True \\
            (191,27) &  24.27$\pm$0.23 &            39.76$\pm$2.41 &    3.60$\pm$1.33 &     36.21$\pm$0.22 &                    True \\
            (191,47) &  24.18$\pm$0.40 &            31.87$\pm$2.31 &    4.34$\pm$1.69 &     31.89$\pm$1.58 &                    True \\
            (192,06) &  24.51$\pm$0.27 &            54.00$\pm$2.60 &    2.52$\pm$0.92 &    206.20$\pm$4.28 &                    True \\
            (192,36) &  24.48$\pm$0.40 &            41.97$\pm$2.46 &    5.16$\pm$1.89 &     35.82$\pm$0.84 &                    True \\
            (192,42) &  24.37$\pm$0.29 &            43.62$\pm$2.47 &    2.14$\pm$0.78 &     24.31$\pm$5.10 &                    True \\
            (193,05) &  24.61$\pm$0.29 &            39.75$\pm$2.41 &    2.75$\pm$0.35 &     90.86$\pm$9.91 &                    True \\
            (193,18) &  24.44$\pm$0.37 &            42.30$\pm$2.51 &   11.27$\pm$4.05 &    206.48$\pm$4.19 &                    True \\
            (193,21) &  24.48$\pm$0.36 &            46.19$\pm$2.49 &    2.90$\pm$0.67 &     73.62$\pm$9.82 &                    True \\
            (193,23) &  23.72$\pm$0.19 &            43.08$\pm$2.46 &    2.82$\pm$0.59 &   173.42$\pm$11.60 &                    True \\
            (193,50) &  24.31$\pm$0.29 &            32.36$\pm$2.51 &   17.95$\pm$7.29 &     42.20$\pm$1.84 &                    True \\
            (301,46) &  23.76$\pm$0.57 &            44.95$\pm$2.61 &    2.02$\pm$0.74 &      8.43$\pm$5.74 &                    True \\
            (302,06) &  22.83$\pm$0.29 &            38.44$\pm$3.69 &  30.55$\pm$17.70 &    199.08$\pm$2.98 &                    True \\
            (195,47) &  22.97$\pm$0.15 &            32.48$\pm$2.45 &    9.90$\pm$3.56 &     14.35$\pm$8.85 &                    True \\
            (202,48) &  23.26$\pm$0.17 &            40.88$\pm$2.58 &   16.22$\pm$5.92 &     47.86$\pm$3.56 &                    True \\
            (203,09) &  24.18$\pm$0.68 &            44.53$\pm$2.48 &    2.08$\pm$0.36 &   166.51$\pm$12.63 &                    True \\
            (203,11) &  23.92$\pm$0.20 &            48.25$\pm$2.90 &   24.39$\pm$9.49 &    215.16$\pm$1.73 &                    True \\
            (203,43) &  24.52$\pm$0.47 &            44.81$\pm$2.53 &    8.68$\pm$3.05 &     45.78$\pm$2.30 &                    True \\
            (205,18) &  24.04$\pm$0.59 &            41.57$\pm$2.47 &    3.22$\pm$0.04 &    122.65$\pm$4.57 &                    True \\
            (284,29) &  22.06$\pm$0.46 &            34.99$\pm$2.54 &  173.36$\pm$0.54 &   136.65$\pm$10.42 &                    True \\
            (285,22) &  20.75$\pm$0.07 &            41.60$\pm$3.94 &   14.00$\pm$6.38 &   349.32$\pm$17.52 &                    True \\
            (296,28) &  22.69$\pm$0.16 &            41.22$\pm$2.71 &    6.75$\pm$1.43 &   252.08$\pm$13.39 &                    True \\
            (300,30) &  23.70$\pm$0.98 &            46.90$\pm$2.94 &   17.79$\pm$7.56 &     24.73$\pm$0.84 &                    True \\
            (017,46) &  23.99$\pm$0.33 &            35.59$\pm$2.45 &    5.31$\pm$0.24 &    285.01$\pm$7.11 &                   False \\
            (018,20) &  23.93$\pm$1.26 &            38.64$\pm$2.45 &    4.92$\pm$0.10 &    316.13$\pm$5.57 &                   False \\
            (018,52) &  24.78$\pm$0.80 &            46.17$\pm$2.57 &   12.08$\pm$3.39 &    243.09$\pm$8.42 &                   False \\
            (191,19) &  24.99$\pm$0.58 &            42.91$\pm$2.44 &    1.58$\pm$0.62 &     39.44$\pm$1.65 &                   False \\
            (191,50) &  24.14$\pm$0.36 &            30.67$\pm$1.17 &  156.06$\pm$4.63 &    216.93$\pm$0.01 &                   False \\
            (192,05) &  24.41$\pm$0.24 &            41.54$\pm$2.43 &    1.07$\pm$0.49 &    58.69$\pm$10.24 &                   False \\
            (192,08) &  24.20$\pm$0.23 &            30.66$\pm$1.17 &  155.99$\pm$4.65 &    216.93$\pm$0.01 &                   False \\
            (192,08) &  24.74$\pm$0.24 &            40.49$\pm$2.42 &    0.80$\pm$0.37 &     42.48$\pm$2.53 &                   False \\
            (192,54) &  22.24$\pm$0.04 &            32.95$\pm$1.16 &  166.54$\pm$2.55 &     41.03$\pm$0.58 &                   False \\
            (193,07) &  24.42$\pm$0.33 &            41.65$\pm$2.44 &    3.00$\pm$0.38 &     91.13$\pm$9.69 &                   False \\
            (193,10) &  24.65$\pm$0.52 &            42.19$\pm$2.44 &    2.70$\pm$0.37 &    89.64$\pm$10.07 &                   False \\
            (193,14) &  23.96$\pm$0.41 &            42.47$\pm$2.45 &    4.01$\pm$1.17 &     62.14$\pm$7.95 &                   False \\
            (193,14) &  25.02$\pm$0.93 &            41.86$\pm$2.43 &    1.83$\pm$0.10 &    107.81$\pm$9.23 &                   False \\
            (193,26) &  23.97$\pm$0.14 &            42.55$\pm$2.44 &    3.06$\pm$0.93 &     63.12$\pm$8.68 &                   False \\
            (193,32) &  24.48$\pm$0.64 &            36.70$\pm$2.37 &    4.86$\pm$1.69 &     50.56$\pm$5.10 &                   False \\
            (193,40) &  24.62$\pm$0.31 &            42.11$\pm$2.44 &    2.22$\pm$0.62 &     66.85$\pm$9.23 &                   False \\
            (301,30) &  22.65$\pm$1.01 &            41.40$\pm$2.73 &   16.18$\pm$6.70 &    203.61$\pm$0.50 &                   False \\
            (301,40) &  23.69$\pm$0.40 &            41.28$\pm$2.65 &    0.63$\pm$0.24 &   238.25$\pm$15.08 &                   False \\
            (305,14) &  23.27$\pm$0.34 &            39.36$\pm$2.42 &    7.04$\pm$2.54 &     36.55$\pm$3.74 &                   False \\
            (305,28) &  23.40$\pm$0.43 &            41.71$\pm$2.44 &    1.24$\pm$0.04 &    129.48$\pm$8.93 &                   False \\
            (305,60) &  23.23$\pm$0.46 &            33.71$\pm$2.69 &  24.33$\pm$10.40 &    206.56$\pm$0.24 &                   False \\
            (306,47) &  23.42$\pm$0.28 &            42.58$\pm$2.51 &    4.02$\pm$0.99 &    63.85$\pm$10.83 &                   False \\
            (306,48) &  22.90$\pm$0.54 &            38.66$\pm$2.59 &   13.04$\pm$5.12 &     37.81$\pm$4.62 &                   False \\
            (306,49) &  22.93$\pm$0.69 &            38.26$\pm$2.46 &    2.75$\pm$0.07 &    130.35$\pm$5.44 &                   False \\
            (307,05) &  23.36$\pm$0.69 &            32.65$\pm$2.60 &    5.20$\pm$0.24 &    135.87$\pm$7.47 &                   False \\
            (310,29) &  22.95$\pm$0.18 &            46.00$\pm$2.53 &    4.40$\pm$0.11 &    105.05$\pm$5.51 &                   False \\
            (310,29) &  23.01$\pm$0.55 &            47.06$\pm$2.54 &    4.18$\pm$0.03 &    125.85$\pm$3.95 &                   False \\
            (310,36) &  23.93$\pm$0.68 &            39.59$\pm$3.05 &  28.70$\pm$12.84 &    202.30$\pm$3.92 &                   False \\
            (311,46) &  22.38$\pm$0.19 &            41.20$\pm$2.90 &  25.12$\pm$10.68 &     23.80$\pm$0.88 &                   False \\
            (313,60) &  22.85$\pm$0.24 &            34.45$\pm$2.43 &    5.22$\pm$1.24 &   248.71$\pm$11.50 &                   False \\
            (314,13) &  23.21$\pm$0.26 &            34.06$\pm$2.38 &    4.65$\pm$1.81 &    213.63$\pm$2.65 &                   False \\
            (316,45) &  23.17$\pm$0.51 &            45.23$\pm$2.49 &    6.53$\pm$2.19 &     14.69$\pm$4.92 &                   False \\
            (316,55) &  23.06$\pm$0.22 &            42.89$\pm$2.46 &    6.75$\pm$2.36 &     19.17$\pm$3.99 &                   False \\
            (194,18) &  24.61$\pm$0.68 &            43.06$\pm$2.45 &    1.51$\pm$0.34 &   351.73$\pm$13.15 &                   False \\
            (194,21) &  24.42$\pm$0.26 &            43.55$\pm$2.46 &    2.94$\pm$0.83 &      5.44$\pm$9.74 &                   False \\
            (194,27) &   24.52$\pm$... &            34.99$\pm$2.46 &   14.26$\pm$5.49 &    222.73$\pm$2.60 &                   False \\
            (195,20) &  24.14$\pm$0.21 &            48.32$\pm$2.78 &   21.31$\pm$7.74 &    225.49$\pm$3.55 &                   False \\
            (195,60) &  24.84$\pm$0.70 &            34.81$\pm$2.41 &    4.34$\pm$0.03 &    311.05$\pm$4.15 &                   False \\
            (196,30) &  24.61$\pm$0.57 &            35.51$\pm$2.58 &   14.99$\pm$6.01 &     25.81$\pm$5.82 &                   False \\
            (197,19) &  24.46$\pm$0.49 &            43.06$\pm$2.46 &    2.46$\pm$0.34 &   341.70$\pm$11.82 &                   False \\
            (197,34) &  22.84$\pm$0.23 &            38.20$\pm$2.45 &   10.77$\pm$3.82 &    229.64$\pm$4.24 &                   False \\
            (197,36) &  24.75$\pm$0.73 &            41.91$\pm$2.45 &    3.26$\pm$0.80 &   359.82$\pm$11.28 &                   False \\
            (197,58) &  24.13$\pm$0.60 &            35.47$\pm$2.71 &   23.38$\pm$9.82 &     32.45$\pm$2.94 &                   False \\
            (202,05) &  23.88$\pm$0.13 &            41.77$\pm$2.45 &    3.54$\pm$0.03 &    121.93$\pm$4.22 &                   False \\
            (202,20) &  23.60$\pm$0.40 &            46.02$\pm$2.53 &    8.61$\pm$2.62 &     58.42$\pm$6.64 &                   False \\
            (202,27) &  23.44$\pm$0.28 &            39.07$\pm$2.44 &    8.62$\pm$2.75 &     57.97$\pm$6.73 &                   False \\
            (202,36) &  23.71$\pm$0.29 &            38.38$\pm$3.22 &  34.18$\pm$15.74 &     42.95$\pm$2.59 &                   False \\
            (202,40) &  23.43$\pm$0.24 &            34.43$\pm$3.42 &  35.99$\pm$18.64 &     41.40$\pm$2.35 &                   False \\
            (202,42) &  24.27$\pm$0.60 &            42.86$\pm$4.42 &  46.16$\pm$26.51 &     41.31$\pm$2.46 &                   False \\
            (203,12) &  24.18$\pm$0.56 &            45.41$\pm$2.56 &   10.71$\pm$3.84 &    209.74$\pm$3.52 &                   False \\
            (203,43) &  24.57$\pm$1.29 &            41.76$\pm$2.45 &    1.04$\pm$0.01 &   130.90$\pm$14.38 &                   False \\
            (204,21) &  23.95$\pm$1.04 &            46.44$\pm$2.54 &    5.71$\pm$2.04 &    211.76$\pm$3.18 &                   False \\
            (204,21) &  24.04$\pm$0.45 &            42.46$\pm$2.48 &    2.51$\pm$0.78 &     62.47$\pm$7.19 &                   False \\
            (204,27) &  24.00$\pm$0.42 &            47.18$\pm$2.59 &   10.79$\pm$3.91 &    216.79$\pm$1.34 &                   False \\
            (204,41) &  24.07$\pm$0.48 &            44.49$\pm$2.50 &    0.61$\pm$0.24 &   177.43$\pm$21.56 &                   False \\
            (205,22) &  24.37$\pm$0.45 &            38.39$\pm$2.49 &    7.67$\pm$2.68 &    198.45$\pm$7.99 &                   False \\
            (205,23) &  24.14$\pm$0.33 &            41.25$\pm$2.53 &   11.44$\pm$4.13 &    205.70$\pm$5.58 &                   False \\
            (205,49) &  24.26$\pm$0.40 &            43.63$\pm$2.49 &    2.33$\pm$0.06 &    139.21$\pm$8.76 &                   False \\
            (284,13) &  23.45$\pm$0.18 &             36.40$\pm$... &    31.17$\pm$... &      10.41$\pm$... &                   False \\
            (284,29) &  23.59$\pm$0.68 &            37.56$\pm$4.53 &    9.23$\pm$4.19 &   340.77$\pm$22.78 &                   False \\
            (284,42) &  22.60$\pm$0.18 &            41.54$\pm$3.52 &   10.56$\pm$4.43 &   238.99$\pm$18.72 &                   False \\
            (284,52) &  23.68$\pm$0.27 &            41.11$\pm$3.67 &   10.45$\pm$3.41 &   340.23$\pm$16.58 &                   False \\
            (284,59) &  23.25$\pm$0.62 &            36.70$\pm$4.97 &    8.47$\pm$3.74 &   254.19$\pm$32.60 &                   False \\
            (288,15) &  23.89$\pm$0.48 &            45.70$\pm$3.47 &   17.81$\pm$9.31 &     12.57$\pm$4.05 &                   False \\
            (288,29) &  23.41$\pm$0.28 &            44.98$\pm$4.41 &  30.20$\pm$20.59 &    204.55$\pm$3.45 &                   False \\
            (288,48) &  23.54$\pm$0.27 &            40.10$\pm$3.98 &  19.55$\pm$13.47 &    208.92$\pm$6.35 &                   False \\
            (289,48) &  23.51$\pm$0.41 &            38.02$\pm$6.91 &  21.02$\pm$27.89 &   213.72$\pm$18.67 &                   False \\
            (289,48) &  23.82$\pm$0.51 &            42.45$\pm$3.30 &    8.96$\pm$3.85 &   233.24$\pm$15.66 &                   False \\
            (290,23) &  23.84$\pm$0.37 &            47.45$\pm$3.38 &  26.35$\pm$12.60 &    211.83$\pm$4.95 &                   False \\
            (291,08) &  23.25$\pm$0.26 &             34.80$\pm$... &    40.90$\pm$... &      18.00$\pm$... &                   False \\
            (291,27) &  23.85$\pm$0.61 &           33.10$\pm$10.91 &  29.45$\pm$64.35 &   206.97$\pm$13.80 &                   False \\
            (296,43) &  23.60$\pm$0.31 &            38.91$\pm$2.78 &    6.00$\pm$0.71 &   327.10$\pm$10.71 &                   False \\
            (297,46) &  23.33$\pm$0.27 &            33.72$\pm$4.03 &    7.33$\pm$0.07 &    291.61$\pm$6.21 &                   False \\
            (298,22) &  23.95$\pm$0.41 &            45.59$\pm$4.09 &  36.73$\pm$20.72 &    216.21$\pm$7.06 &                   False \\
            (298,26) &  22.38$\pm$0.59 &             33.54$\pm$... &    39.53$\pm$... &      17.32$\pm$... &                   False \\
            
\caption{Best-fit parameters estimated from the DECam NEO Survey data for the objects detected with KBMOD. Objects are identified based on their detected pointing group (pg) and CCD. Orbital values and uncertainties are found using the method of \citet{Bernstein2000}. \textit{VR} magnitudes are found as described in \ref{sec:mag}. \textit{VR} magnitude uncertainties are $\sigma_G$ uncertainties, estimated based on the individual stamp \textit{VR} magnitude estimates. Parameters for which no uncertainty was returned are indicated with an ellipsis in the uncertainty value.}

\end{longtable}

\end{document}